\documentclass[fleqn,useAMS,usenatbib]{mn2e}
\usepackage[english]{babel}

\usepackage{fancyhdr}
\usepackage{amsfonts}
\usepackage{amsmath}
\usepackage{amssymb}
\usepackage{multicol}
\usepackage{layout}
\usepackage{graphicx}
\usepackage{epsfig}
\usepackage{verbatim}
\usepackage[T1]{fontenc}
\usepackage{aecompl}
\usepackage{color}
\usepackage{multirow}

\usepackage{times}
\usepackage{natbib}

\newif\ifAMStwofonts
\AMStwofontstrue

\graphicspath{{./fig/}}

\title[Spherical collapse in ghost dark energy]
{Effects of ghost dark energy perturbations on the evolution of spherical overdensities}

\author[Malekjani et~al.]{Mohammad Malekjani$^{1}$ \thanks{malekjani@basu.ac.ir}, Tayebe Naderi$^{1}$ and
Francesco Pace$^{2}$\\
$^1$ Department of Physics, Bu-Ali Sina University, Hamedan 65178, Iran.\\
$^2$ Jodrell Bank Centre for Astrophysics, School of Physics and Astronomy, The University of Manchester, Manchester,
M13 9PL, U.K.}

\begin{document}

\date{Accepted ?, Received ?; in original form \today}

\pagerange{\pageref{firstpage}--\pageref{lastpage}}\pubyear{0}

\maketitle

\label{firstpage}

\begin{abstract}
While in the standard cosmological model the accelerated expansion of the Universe is explained by invoking the
presence of the cosmological constant term, it is still unclear the true origin of this stunning observational fact.
It is therefore interesting to explore alternatives to the simplest scenario, in particular by assuming a more general
framework where the fluid responsible of the accelerated expansion is characterised by a time-dependant equation of
state. Usually these models, dubbed dark energy models, are purely phenomenological, but in this work we concentrate
on a theoretically justified model, the ghost dark energy model. Within the framework of the spherical collapse model,
we evaluate effects of dark energy perturbations both at the linear and non-linear level and transfer these results
into an observable quantity, the mass function, by speculatively taking into account contributions of dark energy to
the mass of the halos. We showed that the growth rate is higher in ghost models and that perturbations enhance the
number of structures with respect to the $\Lambda$CDM model, with stronger effects when the total mass takes into
account dark energy clumps.
\end{abstract}

\section{Introduction}\label{sect:intro}
In the standard cosmological model, the small initial fluctuations
that were seeded during the phase of inflationary expansion, some
$10^{-35}$ seconds after the Big Bang, are the origin of the large
scale structures that we observe today, i.e., galaxies and clusters
of galaxies \citep{Starobinsky1980,Guth1981,Linde1990}. Subsequently
these fluctuations grew under the influence of gravitational
collapse
\citep{Gunn1972,Press1974,White1978,Peebles1993,Sheth1999,Peacock1999,Barkana2001,Peebles2003,Ciardi2005,Bromm2011}.
Most of the growth has taken place after the decoupling of photons
and electrons. The spherical collapse model (SCM) introduced by
\cite{Gunn1972} is a simple analytical model to study the evolution
of the growth of cosmic structures. The scales of interest in the
SCM are much smaller than the Hubble length and the velocities are
non-relativistic. Therefore, pseudo-Newtonian gravity can be used to
study the evolution of the overdensities. With pseudo-Newtonian
gravity we refer to the fact that it is possible to use Newton's
hydrodynamical equations in an expanding Universe including
relativistic contributions to the Poisson equation, i.e. the
contribution of pressure terms. At early times, when the
overdensities are small, linear theory is able to follow the
evolution of spherical overdense regions. In this phase, due to self
gravity, the overdense region expands at a slower rate compared to
the Hubble flow. At a particular scale factor, depending on the
particular background history, the spherical region reaches a
maximum radius and completely detaches from the background
expansion. This is the so called turn-around epoch. Subsequently the
collapse proceeds under the overdense region own gravity and the
system reaches a final steady state with a specific radius due to
virialization processes. The dynamics of the SCM depends strongly on
the evolution of the background Hubble flow. The SCM was extended
and improved in several works
\citep{Fillmore1984,Bertschinger1985,Hoffman1985,Ryden1987,Avila-Reese1998,Subramanian2000,Ascasibar2004,
Williams2004}. More recently, this formalism has been extended to
include shear and rotation terms
\citep{DelPopolo2013a,DelPopolo2013b,DelPopolo2013c} and
non-minimally coupled models \citep{Pace2014b}.

On the other hand, recent developments in observational cosmology
using high quality data including Type Ia supernovae (SNe Ia),
cosmic microwave background (CMB), baryonic acoustic oscillations
(BAO) and Large Scale Structure (LSS), converge to a standard
cosmological model in a spatially flat geometry with a cosmic dark
sector usually in the form of pressureless cold dark matter (CDM)
and dark energy (DE) with negative pressure, respectively, in order
to interpret the observed flat rotation curves of spiral galaxies
and the accelerated expansion of the Universe
\citep{Riess1998,Perlmutter1999,Jaffe2001,Riess2004,Tegmark2004a,Eisenstein2005,Riess2007,Ho2008,Percival2010,
Jarosik2011,Komatsu2011,Planck2013_XIX,Planck2013_XIXX}. On the
basis of the Planck experiment results
\citep{Planck2013_XIX,Planck2013_XIXX}, DE amounts to $\sim68\%$,
CDM and usual baryons to $\sim27\%$ and $\sim5\%$ of the total
energy budget of the Universe, respectively.

Einstein's cosmological constant $\Lambda$ with time independent Equation-of-State (EoS) parameter $w_{\Lambda}=-1$
is the first and simplest candidate to describe DE. Although $\Lambda$ is the simplest model, it suffers from severe
theoretical and conceptual problems: the fine-tuning and the cosmic coincidence problems
\citep{Sahni2000,Weinberg1989,Carroll2001,Peebles2003,Padmanabhan2003,Copeland2006}.
Moreover since $\Lambda$ is constant in space and time, it does not cluster and has a negligible contribution to the
energy density budget of the Universe at high redshifts, so that it affects the evolution of structures in the
Universe at $z\leq 1$.

In order to solve or at least alleviate the theoretical problems of
the $\Lambda$CDM Universe, a wealth of dynamical DE models with a
time varying equation-of-state (EoS) parameter $w_{\rm de}(z)$ has
been proposed. Scalar fields models including quintessence
\citep{Wetterich1988}, phantom
\citep{Caldwell2002,Nojiri2003a,Nojiri2003b}, k-essence
\citep{ArmendarizPicon2000,Chiba2000}, tachyon
\citep{Sen2002,Padmanabhan2002} and dilaton
\citep{Gasperini2002,ArkaniHamed2004,Piazza2004} are few examples of
such dynamical DE models which have been discussed extensively in
literature.

Other dynamical DE models which can successfully interpret the current accelerating Universe are constructed on the
basis of quantum gravity theories. Models such as holographic dark energy (HDE) models
\citep{Horava2000,Thomas2002} and agegraphic dark energy (ADE) models \citep{Cai2007} are derived in the
framework of quantum gravity, by introducing a new degree of freedom or by modifying the theory of gravity
\citep{Horava2000,Thomas2002,Cai2007}. Recently, ghost dark energy models have attracted a lot of
interests in the category of dynamical DE models (see section \ref{sect:GDE} for a in depth description of this
class of models).

Dynamical DE models not only alleviate the theoretical problems affecting the cosmological constant $\Lambda$, but
also, like pressureless matter, possess fluctuations. Hence these models directly affect:\\
i) the dynamics of the background cosmology through the modification of the Hubble parameter,\\
ii) the matter power spectrum and large scale clustering via their fluctuations. \\
Two parameters are important and give a significant influence to DE
perturbations on the matter power spectrum and large scale structure
formation. The first is that the EoS parameter of DE should be
different from $-1$. The second is the effective sound speed $c_{\rm
e}^2$ which connects the density and pressure perturbations
according to $\delta p=c_{\rm e}^2 \delta\rho$. In fact, the sound
horizon of DE is given by $c_{\rm s}H^{-1}$. DE can cluster only on
scales larger than the sound horizon and smaller than the
gravitational horizon, $c_{\rm e}H^{-1}<\lambda<H^{-1}$. Here we
consider two different cases (in units of the speed of light, $c$)
usually discussed in literature: $c_{\rm e}^2=1$ (smooth DE) and
$c_{\rm e}^2=0$ (clustering DE). In the first case, the sound
horizon of DE is equal to the Hubble length, so that DE
perturbations occur on scales equal to or larger than the Hubble
horizon and are therefore negligible on sub-Horizon scales. In the
second scenario, the sound speed is very small compared to the speed
of light $c$. Therefore the sound horizon of DE is significantly
smaller than the Hubble horizon. In this case one can consider DE
perturbations with a wavelength larger than the sound horizon and
smaller than the gravitational Hubble horizon which can grow in a
similar fashion to matter perturbations which growth under
gravitational instability
\citep{Abramo2009a,Appleby2013,Pace2014b,Mehrabi2014}.

The SCM has been investigated and improved in clustering DE models
in order to study how DE perturbations impact structure formation in
the highly non-linear regime
\citep{Abramo2007,Abramo2009b,Pace2014a}. In this work we extend the
SCM in ghost DE models by taking into account the perturbations of
the DE fluid. We study the evolution of overdensities in ghost DE
models and obtain the linear overdensity threshold for collapse
$\delta_{\rm c}$ as well as the virial overdensity $\Delta_{\rm
vir}$ parameters. We show how these quantities are affected by DE
clustering in ghost DE models. We then follow the SCM in clustering
ghost DE Universes by calculating the mass function and the cluster
number count with the Press-Schechter formalism.

The paper is organized as following. In section~\ref{sect:GDE} we introduce the ghost DE models and describe the
evolution of background cosmology in these models.
In section~\ref{sect:SCM} the non-linear SCM in clustering ghost DE models is presented.
In section~\ref{sect:results} we compute the predicted mass function and cluster number count in ghost DE models
using the Press-Schechter formalism by taking into account the clustering DE sector. Finally we conclude and
summarise our results in section~\ref{sect:conclusions}.

\section{Background cosmology in ghost dark energy models}\label{sect:GDE}
The origin of ghost fields traces back to Veneziano ghosts which
have been proposed to find a solution to the $U(1)$ problem in the
low energy effective QCD theory
\citep{Veneziano1979,Witten1979,Kawarabayashi1980,Rosenzweig1980}.
Although in this formalism the ghost field has no contribution in
the flat Minkowski spacetime, in curved spacetime it has a small
energy density proportional to $\Lambda^3_{\rm QCD}H$, where $H$ is
the Hubble parameter and $\Lambda^3_{\rm QCD}$ is the QCD mass scale
\citep{Ohta2011}. With $\Lambda^3_{\rm QCD}\sim 100$ MeV and $H\sim
10^{-33}$~eV, the quantity $\Lambda^3_{\rm QCD}H$ gives the right
order of magnitude of the observed energy density of DE
$(\sim3\times10^{-3}~{\rm eV})^4$ \citep{Ohta2011}. This small
vacuum energy density can be considered as a driver engine for the
evolution of the Universe. Comparing with other theoretical
dynamical DE models like HDE and ADE models in which one should
introduce a new parameter or a new degree of freedom, the most
important advantage of ghost DE models is that they come from the
standard model of particle physics and are totally embedded both in
the standard model and in general relativity without introducing any
new parameter or new degrees of freedom. This numerical coincidence
also shows that this model can solve the fine tuning problem
\citep{Urban2009a,Urban2009b,Urban2010a,Urban2010b}. As it was
mentioned above, the energy density in ghost DE model is
proportional to the Hubble parameter \citep{Ohta2011}:
\begin{equation}\label{eqn:GDE1}
 \rho_{\rm de}=\alpha H\;,
\end{equation}
where $\alpha>0$ is roughly of order $\Lambda^3_{\rm QCD}$. From the
observational point of view, ghost DE models can fit a whole set of
cosmological data including SNe Ia, BAO, CMB, Big Bang
nucleosynthesis (BBN) and Hubble parameter data \citep{Cai2011a}.
The cosmological evolution of the models has been investigated in
\citep{Urban2009a,Urban2009b,Urban2010a,Urban2010b} who stated that
the Universe begins to accelerate at redshift around $z\sim0.6$.

The Friedmann equation for a Universe containing pressureless dust matter and DE in a flat geometry described by a
Friedmann-Robertson-Walker (FRW) metric is given by
\begin{equation}\label{eqn:fh3}
 H^{2}=\frac{8\pi G}{3}(\rho_{\rm m}+\rho_{\rm de})\;,
\end{equation}
where $\rho_{\rm m}$ and $\rho_{\rm de}$ are the energy density of the pressureless dust matter and DE components,
respectively, and $H$ is the Hubble parameter. We use equation~(\ref{eqn:GDE1}) for the energy density of the DE
component and insert it into equation~(\ref{eqn:fh3}) in order to obtain the Hubble parameter in ghost DE cosmologies
\begin{equation}\label{eqn:Hubb1}
 H=\frac{4\pi G}{3}\alpha+\sqrt{\left(\frac{4\pi G}{3}\alpha\right)^2+\frac{8\pi G}{3}\rho_{\rm m0}a^{-3}}\;.
\end{equation}
In terms of the dimensionless energy density $\Omega_{\rm m,0}=8\pi G\rho_{\rm m,0}/(3H_{0}^2)$ and redshift parameter
$z=1/a-1$, the above Hubble equation becomes
\begin{equation}\label{eqn:Hubb2}
H(z)=H_0\left(\kappa+\sqrt{\kappa^2+\Omega_{\rm m,0}(1+z)^3}\right)\;,
\end{equation}
where $\kappa=(1-\Omega_{\rm m,0})/2$.

In the above mentioned studies, the energy density of ghost DE is
assumed to be proportional to the Hubble parameter via
equation~(\ref{eqn:GDE1}). However, the energy density of the
Veneziano ghost field in QCD theory is generally of the form
$H+\emph{O}(H^2)$ \citep{Zhitnitsky2011}. Although in ghost DE
models only the leading term $H$ was assumed, it has been shown that
the sub-leading term $H^2$ can also be important in the early
evolution of the Universe \citep{Maggiore2012}. A ghost DE model
with sub-leading term $H^2$ is usually called generalised ghost dark
energy model. \cite{Cai2012} showed that generalised ghost DE models
result in better agreement with observations compared to ordinary
ghost DE models. The energy density of a generalised ghost DE model
is given by \citep{Cai2011a}
\begin{equation}\label{eqn:GGDE1}
 \rho_{\rm de}=\alpha H+\beta H^2\;,
\end{equation}
where $\alpha$ and $\beta$ are the constants of the model. Inserting equation~(\ref{eqn:GGDE1}) into
equation~(\ref{eqn:fh3}), the Hubble parameter becomes
\begin{equation}\label{eqn:Hubb3}
 H=\frac{4\pi G}{3\gamma}\alpha+\sqrt{\left(\frac{4\pi G}{3\gamma}\alpha\right)^2+
 \frac{8\pi G}{3\gamma}\rho_{\rm m0}a^{-3}}\;,
\end{equation}
where $\gamma=1-8\pi G\beta/3$. Setting $\gamma=1$, generalised ghost DE model reduces
to ordinary ghost DE model ($\beta=0$), as expected. In terms of the dimensionless matter energy density $\Omega_{\rm m,0}$ and redshift
parameter $z$, the Hubble equation in generalised ghost DE models is given by
\begin{equation}\label{eqn:Hubb4}
 H(z)=H_0\left(\kappa+\sqrt{\kappa^2+\frac{\Omega_{\rm m,0}(1+z)^3}{\gamma}}\right)\;,
\end{equation}
where $\kappa=(1-\Omega_{\rm m,0}/\gamma)/2$.\\
The conservation equations for pressureless dust matter and DE at the background level are
\begin{eqnarray}
 \dot{\rho}_{\rm m}+3H\rho_{\rm m} & = & 0\;, \label{eqn:contmt}\\
 \dot{\rho}_{\rm de}+3H(1+w_{\rm de})\rho_{\rm de} & = & 0\;,\label{eqn:contdt}
\end{eqnarray}
where the dot is the derivative with respect to cosmic time and $w_{\rm de}$ is the DE EoS parameter.\\
Taking the time derivative of Friedmann equation~(\ref{eqn:fh3}) and using equations~(\ref{eqn:contmt})
and~(\ref{eqn:contdt}) as well as the critical density $\rho_c=3H^2/(8\pi G)$, we obtain
\begin{equation}\label{eqn:hdot}
 \frac{\dot{H}}{H^2}=-\frac{3}{2}(1+\Omega_{\rm de}w_{\rm de})\;,
\end{equation}
where $\Omega_{\rm de}$ is the dimensionless density parameter of the DE component. Differentiating
equations~(\ref{eqn:GDE1}) and~(\ref{eqn:GGDE1}) with respect to time and inserting the results in the conservation
equation for DE (equation~\ref{eqn:contdt}) and also using equation~(\ref{eqn:hdot}), the EoS parameter for ghost DE
and generalised ghost DE models are
\begin{eqnarray}
 w_{\rm de}(z) & = & \frac{1}{\Omega_{\rm de}(z)-2}\;,\label{eqn:eos1}\\
 w_{\rm de}(z) & = & \frac{1-\Omega_{\rm de}(z)-\gamma}{\Omega_{\rm de}(z)\left(1-\Omega_{\rm de}(z)+\gamma\right)}\;.
 \label{eqn:eos2}
 \end{eqnarray}
Setting $\gamma=1$, equation~(\ref{eqn:eos2}) reduces to~(\ref{eqn:eos1}) as expected. We now calculate the equation
of motion for the energy density of DE in ghost DE and generalised ghost DE models. Taking the time derivative of
$\Omega_{\rm de}=\rho_{\rm de}/\rho_{\rm c}$ and using equations~(\ref{eqn:GDE1}) and~(\ref{eqn:hdot}) and finally
changing the time derivative to a derivative with respect to cosmic redshift $z$, we have
\begin{equation}\label{eqn:motion}
\frac{d\Omega_{\rm de}(z)}{dz}=-\frac{3\Omega_{\rm de}(z)}{2(1+z)}\left[1+\Omega_{\rm de}(z)w_{\rm de}(z)\right]\;,
\end{equation}
for both ghost and generalised ghost DE cosmologies. We see that the evolution of the DE density in ghost DE
cosmologies via equation~(\ref{eqn:motion}) depends on the EoS parameter of the models according to the
relations~(\ref{eqn:eos1}) and~(\ref{eqn:eos2}). We solve the system of coupled
equations~(\ref{eqn:Hubb2}),~(\ref{eqn:eos1}),~(\ref{eqn:motion})
and~(\ref{eqn:Hubb4}),~(\ref{eqn:eos2}),~(\ref{eqn:motion}) in order to calculate the evolution of the Hubble and EoS
parameters as well as the energy density of DE in ghost and generalised ghost DE cosmologies, respectively.
To fix the cosmology, the present values of matter density and DE density parameters are chosen as:
$\Omega_{\rm m,0}=0.27$ and $\Omega_{\rm de,0}=0.73$ in a spatially flat Universe. The present Hubble parameter is
$H_0=70$~km/s/Mpc.

\begin{figure}
 \centering
 \includegraphics[width=7cm]{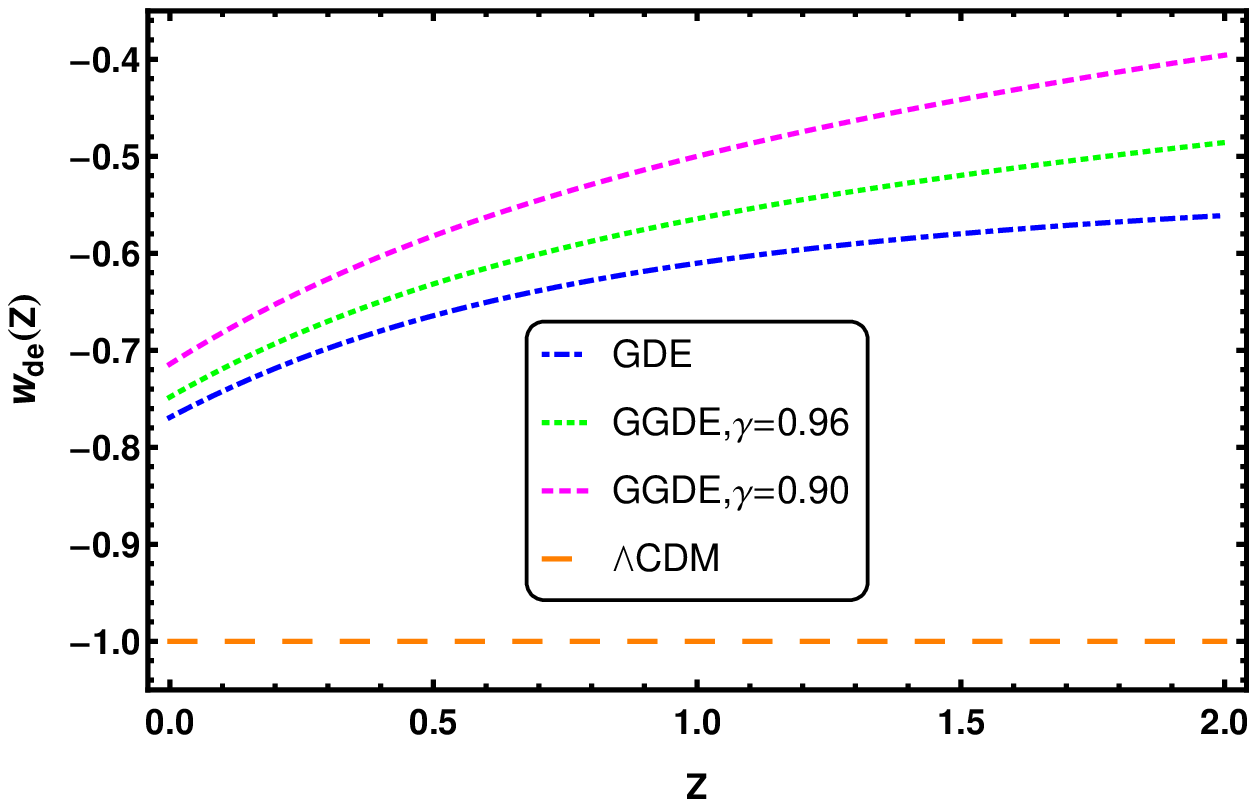}
 \includegraphics[width=7cm]{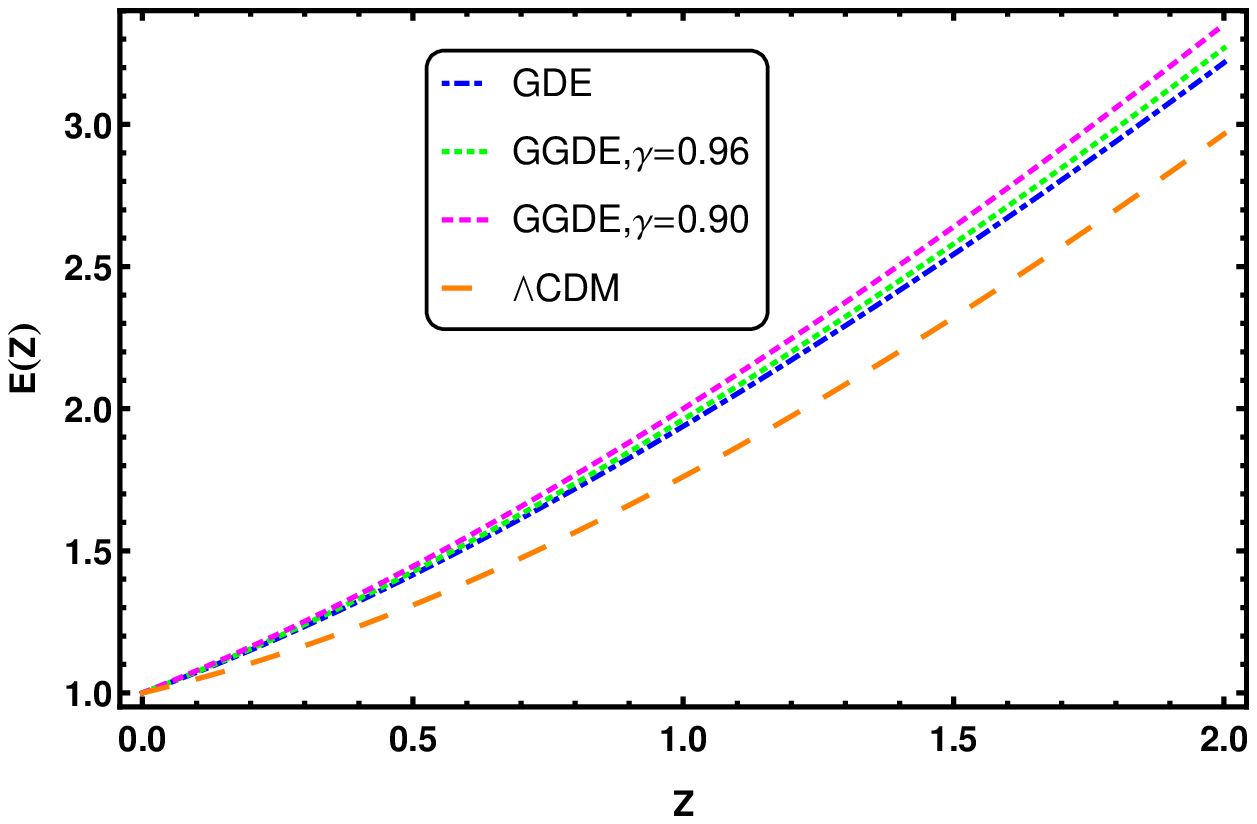}
 \includegraphics[width=7cm]{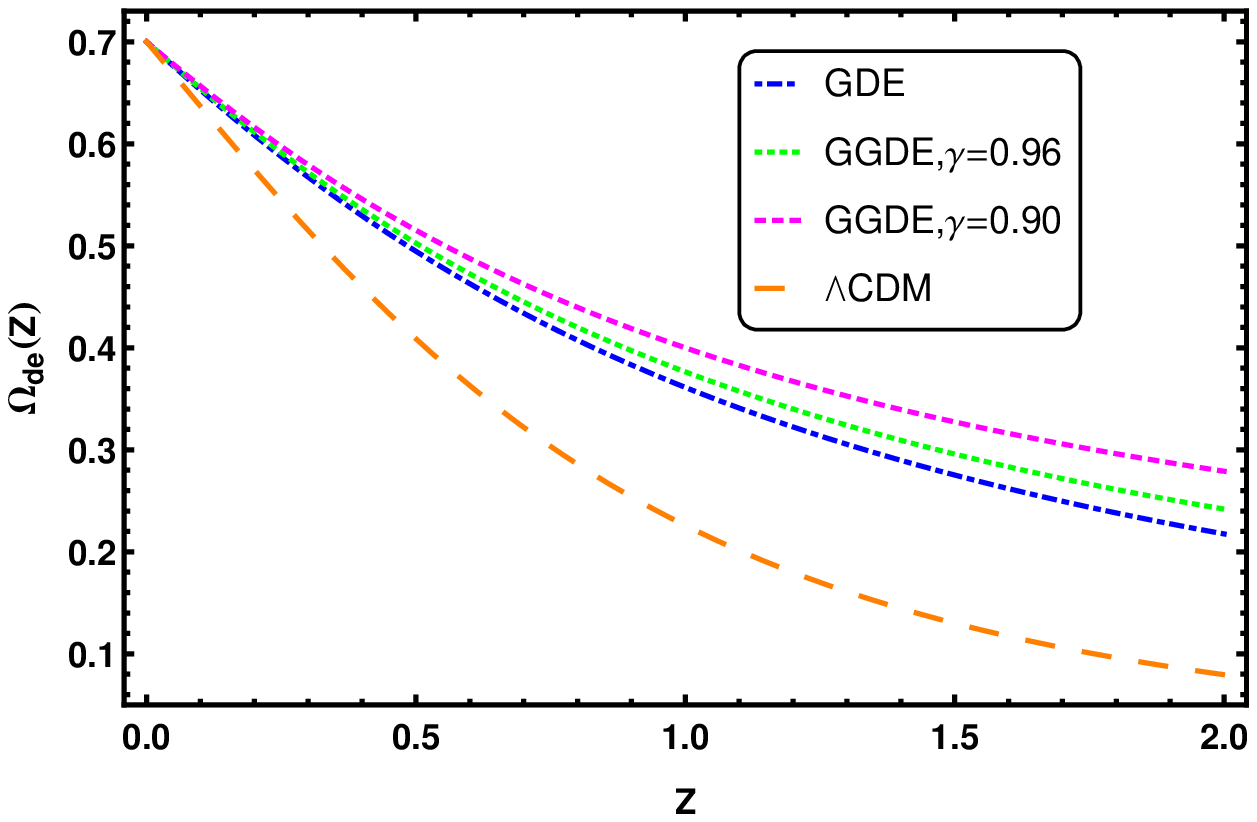}
 \caption{Top panel: Evolution of the EoS parameter $w_{\rm de}$. Middle panel: dimensionless Hubble parameter $E(z)$.
 Bottom panel: DE density parameter $\Omega_{\rm de}$ as a function of cosmic redshift $z$ for the different
 cosmological models considered in this work. As shown in the legend, the ghost DE is indicated by GDE, generalised
 ghost DE by GGDE. The  $\Lambda$CDM  model is shown with the orange dashed line, the GDE model with the red
 short-short dashed line and the  GGDE model with the brown (pink) short-dashed line for $\gamma=0.96$
 ($\gamma=0.90$), respectively.}
 \label{fig:back}
\end{figure}

In figure~(\ref{fig:back}) we show the evolution of the EoS parameter $w_{\rm de}$ (top panel), dimensionless Hubble
parameter $E=H/H_0$ (middle panel) and energy density of DE component $\Omega_{\rm de}$ (bottom panel) as a function
of the cosmic redshift $z$ for different ghost, generalised ghost DE and $\Lambda$CDM cosmological models. In the case
of the generalised ghost DE model we choose the model parameter $\gamma$ as $0.96$ and $0.90$. In all the figures of
this work we indicate the ghost DE as GDE and the generalised ghost DE as GGDE. We see that the EoS parameter for
ghost and generalised ghost DE models is always bigger than $w_{\rm \Lambda}=-1$ and remains in the quintessence
regime, i.e., $w_{\rm de}>-1$. The Hubble parameter and the DE density are bigger in these models compared to the
concordance $\Lambda$CDM Universe. This means that the rate of acceleration is stronger for (generalised) ghost models
and that the relevance of the DE fluid is important for a longer period of the cosmic history.

\section{Spherical collapse in ghost dark energy cosmologies }\label{sect:SCM}
In this section we investigate the SCM in the framework of
clustering ghost and generalised ghost DE cosmologies. The effects
of DE perturbations on the evolution of matter overdensities have
been extensively investigated in literature \citep[see,
i.e.][]{Mota2004,Abramo2007,Abramo2008,Abramo2009a,Creminelli2010,Basse2011,Batista2013,Pace2014a}.
Following \cite{Pace2014a}, the fully perturbed equations for the
evolution of non-relativistic dust matter $\delta_{\rm m}$ and dark
energy $\delta_{\rm de}$ perturbations in the non-linear regime
(without the contribution of shear and rotation) are given by
\begin{eqnarray}
 \delta_{\rm m}^{\prime}+\left(1+\delta_{\rm m}\right)\frac{\tilde{\theta}}{a} & = & 0\;,\label{eqn:DMpt1}\\
 \delta_{\rm de}^{\prime}-\frac{3}{a}w_{\rm de}\delta_{\rm de}+
 \left[1+w_{\rm de}+\delta_{\rm de}\right]\frac{\tilde{\theta}}{a} & = & 0\;,\label{eqn:DEpt1}\\
 \tilde{\theta}^{\prime}+\left(\frac{2}{a}+\frac{E^{\prime}}{E}\right)\tilde{\theta}+
 \frac{\tilde{\theta}^2}{3a}+
 \frac{3}{2a}\left[\Omega_{\rm DM}\delta_{\rm m}+\Omega_{\rm de}\delta_{\rm de}\right]&=&0\; \label{eqn:theta1}
\end{eqnarray}
where $\tilde{\theta}=\theta/H$ is the dimensionless divergence of the comoving peculiar velocity for dust matter and
DE.

The linear evolution of overdensities at early times is
\begin{eqnarray}
\delta_{\rm m}^{\prime}+\frac{\tilde{\theta}}{a}& = & 0\;,\label{eqn:LDMpt1}\\
\delta_{\rm de}^{\prime}-\frac{3}{a}w_{\rm de}\delta_{\rm de}+
\left[1+w_{\rm de}\right]\frac{\tilde{\theta}}{a} & = & 0\;,\label{eqn:LDEpt1}\\
\tilde{\theta}^{\prime}+\left(\frac{2}{a}+\frac{E^{\prime}}{E}\right)\tilde{\theta}+
\frac{3}{2a}\left[\Omega_{\rm DM}\delta_{\rm m}+\Omega_{\rm de}\delta_{\rm de}\right]&=&0\; \label{eqn:Ltheta1}
\end{eqnarray}

As in \cite{Pace2014a}, to determine the initial conditions to solve
the above differential equations describing the evolution of
perturbations, we find the initial value $\delta_{\rm m,i}$ such
that at the collapse scale factor $a_{\rm c}$ the matter overdensity
diverges, $\delta_{\rm m}\rightarrow\infty$. The initial values for
the DE overdensity $\delta_{\rm de,i}$ and peculiar velocity
perturbation $\tilde{\theta_i}$ are related to $\delta_{\rm m,i}$
via \citep{Batista2013,Pace2014a}:
\begin{eqnarray}
 \delta_{\rm de,i} & = & \frac{n}{(n-3w)}(1+w_{\rm de})\delta_{\rm m,i}\;,\label{eqn:deltadei}\\
 \tilde{\theta}_{i} & = & -n\delta_{\rm m,i}\;.\label{eqn:thetaDMi}
\end{eqnarray}
In the case of an EdS model $n=1$. However, in DE cosmologies it has
been shown that there is a small deviation from unity
\citep{Batista2013}. In the limiting case of non-clustering DE
models ($c_{\rm e}=1$) the coupled non-linear equations
(\ref{eqn:DMpt1}, \ref{eqn:DEpt1}, \ref{eqn:theta1}) and system of
linear equations (\ref{eqn:LDMpt1}, \ref{eqn:LDEpt1},
\ref{eqn:Ltheta1}), respectively, reduce to the following equations:
\begin{eqnarray}
\delta_{\rm m}^{\prime\prime}+\left(\frac{3}{a}+\frac{E^{\prime}}{E}\right)\delta_{\rm m}^{\prime}
-\frac{4}{3}\frac{\delta_{\rm m}^{\prime 2}}{1+\delta_{\rm m}}-
\frac{3\Omega_{\rm m0}}{2a^5E^2}\delta_{\rm m}(1+\delta_{\rm m})=0\;, \label{eqn:Nclusteing1}\\
\delta_{\rm m}^{\prime\prime}+\left(\frac{3}{a}+\frac{E^{\prime}}{E}\right)\delta_{\rm m}^{\prime}-
\frac{3\Omega_{\rm m0}}{2a^5E^2}\delta_{\rm m}=0\;, \label{eqn:Nclusteing2}
\end{eqnarray}
as expected \citep[see also][]{Pace2010}. In the case of non-clustering ghost and generalised ghost DE models, we will
solve equations (\ref{eqn:Nclusteing1}, \ref{eqn:Nclusteing2}), instead of the previous ones.

\section{Results}\label{sect:results}
We first solve the system of equations (\ref{eqn:LDMpt1}), (\ref{eqn:LDEpt1}), (\ref{eqn:Ltheta1}) for the case of
clustering DE ($c_{\rm e}=0$) and (\ref{eqn:Nclusteing2}) for non-clustering DE ($c_{\rm e}=1$) in order to
determine the linear evolution of the overdensities $\delta_{\rm m}$ and $\delta_{\rm de}$ ($\delta_{\rm m}$ in the
non-clustering DE case) and obtain the linear growth factor $D_{+}(z)=\delta_{\rm m}(z)/\delta_{\rm m}(z=0)$ as a
function of cosmic redshift $z$ for the different cosmological models considered in this work.

\begin{figure}
 \centering
 \includegraphics[width=8cm]{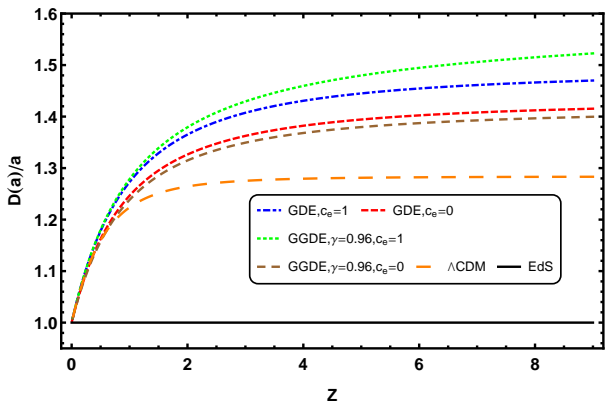}
 \caption{Time evolution of the growth factor normalised to the present time and divided by the scale factor $a$ as a
 function of the cosmic redshift $z$ for different cosmological models. Black solid line shows the EdS model, the
 orange long-dashed line the $\Lambda$CDM model, the blue (green) dot-dashed (dotted) line the homogeneous ghost
 (generalised ghost) DE model, while the red (brown) dashed line the clustering ghost (generalised ghost) DE
 cosmology.}
 \label{fig:gf}
\end{figure}

In figure~(\ref{fig:gf}), the evolution of the growth factor normalised at $z=0$ and divided by the scale factor $a$
is presented as a function of redshift. In the EdS model (black solid line) $D_{+}/a$ is equal to $1$ at any time
showing that the growth of matter perturbations $\delta_{\rm m}$ is the same at all redshifts. In a $\Lambda$CDM
Universe (orange long-dashed curve) the growth factor is higher than the EdS model throughout its history, but falls
for lower redshifts. The lowering of the growth factor at lower redshifts is due to the fact that at late times the
cosmological constant dominates the energy budget of the Universe and suppresses the amplitude of perturbations. On
the other hand, a larger growth factor in a $\Lambda$CDM Universe at higher redshift shows that the growth of dust
matter perturbations will be stronger than in a EdS Universe at early times. For ghost and generalised ghost DE models
we show $D_{+}/a$ for two extreme cases: fully clustering DE ($c_{\rm e}=0$) and homogeneous DE ($c_{\rm e} = 1$)
models. In the case of generalised ghost DE model, we choose a value of $0.96$ for the model parameter $\gamma$. The
red and brown dashed curves stand for clustering ghost and generalised ghost DE models, respectively. The blue
dotted-dashed and green dotted curves show the variation of $D_{+}/a$ for homogeneous ghost and generalised ghost DE
Universes, respectively. We see that the growth factor is largest (intermediate) for homogeneous (clustering) ghost
and generalised ghost DE models compared to the concordance $\Lambda$CDM Universe. It is worth to mention that
$D_{+}/a$ changes more rapidly at low redshifts where dark energy dominates the energy budget of the Universe.
At early times instead, the relative change is much shallower. Differences are obviously still noticeable in the
models. In particular we notice that while $D_{+}/a$ for the $\Lambda$CDM model is almost a constant, this is not
entirely the case for the dark energy models analysed in this work. This is easily interpreted taking into account
the stronger importace of dark energy at early times, as clearly shown in the lower panel of figure~(\ref{fig:back}).
Similarly to the $\Lambda$CDM Universe, in ghost and generalised ghost DE models, DE suppresses the growth of
perturbations at low redshifts. More quantitatively, at high redshifts, the growth of perturbations parameter
$D_{+}/a$ in non-clustering (clustering) generalised ghost DE model is $\approx 18.7\%$ ($\approx10.9\%$) larger than
in the $\Lambda$CDM concordance model. These values are only midly larger than the ghost dark energy models, where
$D_{+}/a$ differs of $\approx14.5\%$ ($\approx10\%$) for non-clustering (clustering) ghost DE models with respect to a
$\Lambda$CDM model.

\subsection{Spherical Collapse Model parameters}
Here we evaluate the two characterising quantities of the spherical collapse model: the linear overdensity parameter
$\delta_{\rm c}$ and the virial overdensity $\Delta_{\rm vir}$ for ghost and generalised ghost DE cosmologies.
The linear overdensity $\delta_{\rm c}$ together with the growth factor $D_{+}$ is important in order to evaluate the
mass function in the Press Schechter formalism \citep{Press1974,Bond1991,Sheth2002}. The virial overdensity
$\Delta_{\rm vir}$ is used to calculate the size of spherically symmetric halos with given mass $M$.

We search the initial conditions $\delta_{\rm m,i}$ following the general approach in \cite{Pace2010,Pace2012} and
then we use equations~(\ref{eqn:deltadei}) and~(\ref{eqn:thetaDMi}) to determine $\delta_{\rm de,i}$ and
$\tilde{\theta_i}$. Once the initial conditions are found, we then solve the system of
equations~(\ref{eqn:LDMpt1}),~(\ref{eqn:LDEpt1}) and~(\ref{eqn:Ltheta1}) in order to obtain
$\delta_{\rm c}=\delta_{\rm m}(z=z_c)$. In the non-clustering case, we solve equation (\ref{eqn:Nclusteing2}) to
calculate $\delta_{\rm c}$.

The collapse redshift $z_{\rm c}$ is defined as the redshift at
which the matter overdensity tends to infinity $\delta_{\rm
m}\rightarrow\infty$ \citep[see
also][]{Pace2010,Pace2012,Pace2014a}.

\begin{figure}
 \centering
 \includegraphics[width=8cm]{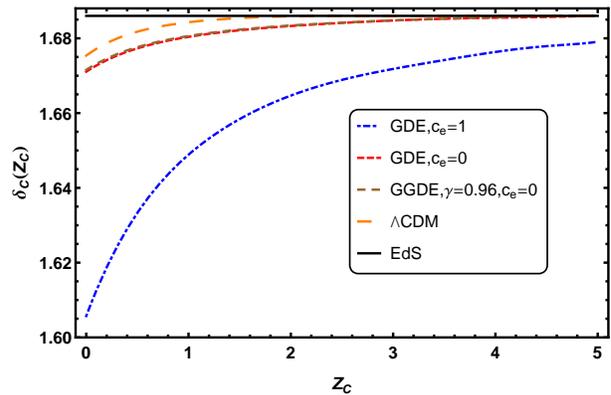}
 \caption{The variation of the linear threshold density contrast $\delta_{\rm c}$ as a function of the collapse
 redshift for the different cosmological models analysed in this work. Line styles and colours are as in
 Fig.~\ref{fig:gf}.}
 \label{fig:deltac}
\end{figure}

In figure~(\ref{fig:deltac}) the evolution of the linear overdensity $\delta_{\rm c}$ as a function of the collapse
redshift $z_{\rm c}$ is presented for different models: EdS, $\Lambda$CDM, homogeneous and clustering ghost DE and
clustering generalised ghost DE. We refer to the caption for line styles and colours of each model. Analogously to the
previous section, we use $\gamma=0.96$ in the case of generalised ghost DE models. In all DE models, the linear
overdensity parameter approaches the fiducial value in the EdS Universe $\delta_{\rm c}\approx 1.686$ at high enough
redshifts, as expected. In fact at high redshifts, the Universe is dominated by pressureless dust matter and the
effects of DE on the scenario of structure formation is negligible. At lower redshifts, $\delta_{\rm c}$ decreases and
deviates from the EdS limit. The important point to note is that in clustering ghost and generalised ghost DE models,
the behaviour of $\delta_{\rm c}$ is more similar to the $\Lambda$CDM Universe with respect to the homogeneous models.
This result is in agreement with what found by \cite{Pace2010} and \cite{Batista2013} for early dark energy models and
\cite{Pace2014b} for clustering quintessence and phantom DE models. Quantitatively we see that the differences
between homogeneous ghost and generalised ghost DE models with the $\Lambda$CDM model at the present time are
roughly $4.5\%$ and $21.5\%$, respectively, while in the case of clustering ghost and generalised ghost models the
differences are very small, of the order of $\approx 0.25\%$ for both cases. As for the case of perturbations in
early dark energy models, we can explain the fact that perturbations in the dark energy fluid make the model more
similar to the $\Lambda$CDM model by taking into account that the Poisson equation is now modified and the
gravitational potential is sourced also by DE perturbations.

\begin{figure}
 \centering
 \includegraphics[width=8cm]{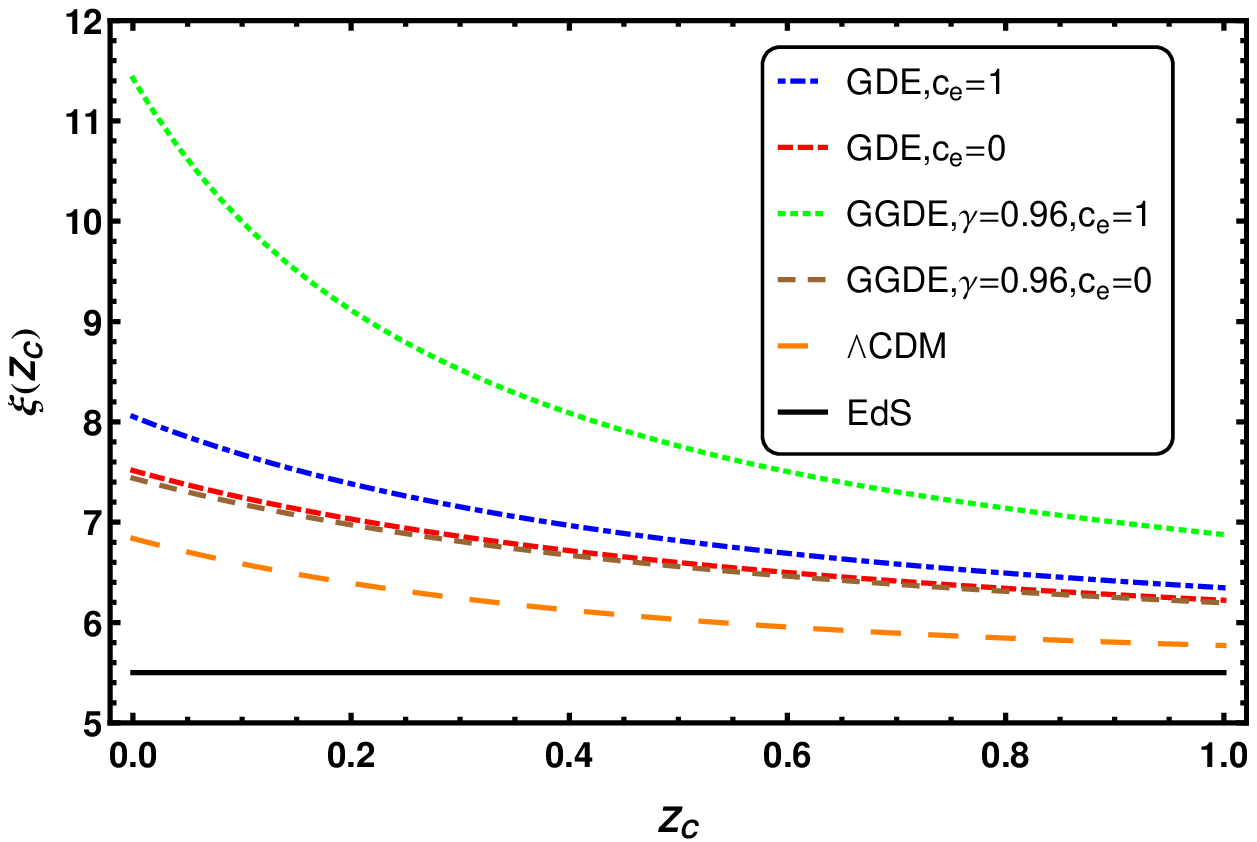}
 \includegraphics[width=8cm]{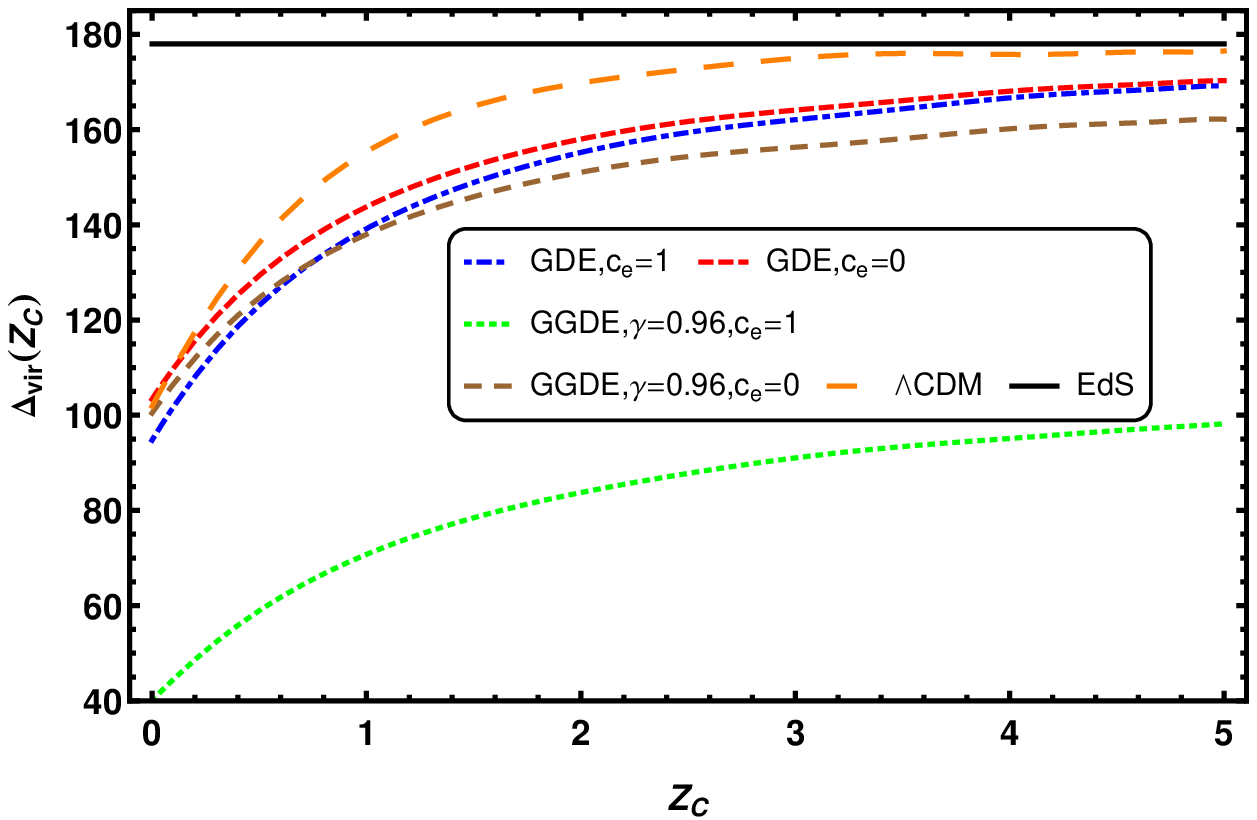}
 \caption{The variation of turn around overdensity $\zeta$ (top panel) and virial overdensity $\Delta_{\rm vir}$
 (bottom panel) with respect to the collapse redshift $z_{\rm c}$ for various models considered in this work. Line
 styles and colours are as in Fig.~\ref{fig:gf}.}
 \label{fig:xi_DeltaV}
\end{figure}

In the framework of the SCM, the virialization process of
pressureless dust matter and the size of forming halos are strongly
affected by the DE sector
\citep{Lahav1991,Wang1998,Mota2004,Horellou2005,Wang2005} and also
its perturbations
\citep{Abramo2007,Abramo2008,Abramo2009a,Batista2013,Pace2014a}. The
virial overdensity is defined as $\Delta_{\rm vir}=\zeta (x/y)^3$,
where $\zeta$ is the overdensity at the turn-around epoch, $x$ is
the scale factor normalised to the turn-around scale factor and $y$
is the ratio between the virialization radius and the turn-around
radius \citep{Wang1998}. In an EdS cosmology, it is simple to show
that $y=1/2$, $\zeta\approx 5.6$ and $\Delta_{\rm vir}\approx178$
independently of the cosmic redshift \citep[see also][]{Naderi2015}.
In DE cosmologies, $\Delta_{\rm vir}$ depends on the evolution of
the DE fluid and it is a redshift dependent quantity. In particular,
according to whether DE takes part or not into the virialization
process, the quantity $y$ may be larger or smaller than $1/2$ and
the parameter $\Delta_{\rm vir}$ can be affected by the clustering
of DE \citep{Maor2005,Pace2014a}.

In the line of these studies, we calculate the turn-around and virial overdensities $\zeta$ and $\Delta_{\rm vir}$ in
ghost and generalised ghost DE cosmologies. We also determine how the clustering of the DE component can change the
variation of $\zeta$ and $\Delta_{\rm vir}$ in these models. Our results are presented in
figure~(\ref{fig:xi_DeltaV}). In the top panel, the variation of the overdensity at the turn-around redshift $\zeta$
is shown for the different DE models investigated in this work. In the limiting case of the EdS model, $\zeta=5.6$
independently of the cosmic time. At early times, $\zeta$ tends to the critical value $\zeta=5.6$ representing the
early matter-dominated era. The value of $\zeta$ is larger for both clustering and non-clustering versions of ghost
and generalised ghost DE models, compared to the concordance $\Lambda$CDM model. Differences between the dark energy
models and the $\Lambda$CDM model cover a relatively huge spectrum of values, according to whether we take into
account the perturbations of the dark energy fluid. In particular for homogeneous (generalised) ghost DE models,
differences are of the order of ($67\%$) $17\%$ while when perturbations are taken into account, differences drop down
to ($9\%$) $10\%$. This shows how the generalised ghost model, is much more sensitive to the inclusion of
DE perturbations. Once again, this is due the higher amount of DE density at earlier time for these kind of models.
Hence we conclude that in the context of ghost DE scenarios the perturbed spherical perturbations detach from the
background Hubble flow with higher overdensities compared to the EdS and $\Lambda$CDM Universes. We also notice that
the results for clustering ghost and generalised ghost DE models are closer to what is found for the reference
$\Lambda$CDM model compared to the non-clustering case.

In the bottom panel, the results for the virial overdensity
$\Delta_{\rm vir}$ are presented. In the case of ghost DE models,
the results are closer to the reference $\Lambda$CDM model. In these
models, the difference between homogeneous and clustering DE is very
small. However, in the case of generalised ghost cosmology the
differences are more pronounced. Differences between ghost DE models
and $\Lambda$CDM model at the present time are of the order of
$60\%$ for non-clustering generalised ghost DE model and $7\%$ for
homogenous ghost DE models, roughly as for the overdensity at the
turn-around radius. This is not surprising, since this quantity is
the main ingredient used to evaluate the virial overdensity. When
perturbations in dark energy are taken into account differences are
only of the order of $1.5\%$ for both classes analysed. These
results are similar to what found in \cite{DelPopolo2013b} and
\cite{Pace2014b}.

\subsection{Mass function and halo number density}
We know that galaxies and cluster of galaxies are embedded in the extended halos of cold dark matter (CDM). In the
Press \& Schechter formalism \citep{Press1974}, the abundance of CDM halos can be described as a function of their
mass and a Gaussian distribution function expresses the fraction of the volume of the Universe which collapses into an
object of mass $M$ at a certain redshift $z$. In this formalism, the comoving number density of virialised structures
with masses in the range of $M$ and $M+dM$ at redshift $z$ is given by
\begin{eqnarray}\label{eqn:PS1}
\frac{dn(M,z)}{dM}=-\frac{\rho_{\rm m0}}{M}\frac{d\ln{\sigma(M,z)}}{dM}f(\sigma)\;,
\end{eqnarray}
where $\sigma$ is the r.m.s. of the mass fluctuation in spheres of mass $M$ and $f(\sigma)$ is the mass function. The
standard mass function in the Press \& Schechter formalism is given by \citep{Press1974}
\begin{equation}\label{eqn:PS2}
f(\sigma)=\sqrt{\frac{2}{\pi}}\frac{\delta_{\rm c}(z)}{\sigma(M,z)}
\exp{\left[-\frac{\delta_{\rm c}^2(z)}{2\sigma^2(M,z)}\right]}\;.
\end{equation}

Although the standard mass function provides a good general representation of the predicted number density of CDM
halos, it deviates from simulations due to an over-prediction of low-mass objects and an under-prediction of high-mass
objects at the present time \citep{Sheth1999,Sheth2002}. Here we use another popular fitting formula proposed by Sheth
\& Tormen \citep{Sheth1999,Sheth2002}, the so-called ST mass function:
\begin{equation}\label{eq:multiplicity_ST}
 f_{\rm ST}(\sigma)= A\sqrt{\frac{2a}{\pi}}\left[1+\left(\frac{\sigma^2(M,z)}{a\delta_{\rm  c}^2(z)}\right)^p\right]
 \frac{\delta_c(z)}{\sigma(M,z)} \exp{\left(-\frac{a\delta_{\rm c}^2(z)}{2\sigma^2(M,z)}\right)}\;,
\end{equation}
where the numerical parameters are: $A=0.3222$, $a=0.707$ and
$p=0.3$. Putting $A=1/2$, $a=1$ and $p=0$, the well known
Press-Schechter mass function can be recovered as expected. In a
Gaussian density field, the mass variance $\sigma^2$ is given by
\begin{equation}\label{eq:sigma}
 \sigma^2=\frac{1}{2\pi^2}\int_0^{\infty}{k^2P(k)}W^2(kR)dk\;,
\end{equation}
where $R=(3M/4\pi\rho_{m,0})^{1/3}$ is the radius of the overdense spherical path at the present time,
$W(kR)=3[\sin(kR)-kR\cos(kR)]/(kR)^3$ is the Fourier transform of a spherical top-hat profile window function with
radius $R$ and $P(k)$ is the linear power spectrum of density fluctuations \citep{Peebles1993}.

The number density of objects above a given mass at a certain fixed redshift is
\begin{equation}\label{eq:ndensity}
 n(>M)=\int_{\rm M}^{\rm \infty}\frac{dn}{dM^{\prime}}dM^{\prime}\;.
\end{equation}

\begin{figure*}
 \begin{center}
  \includegraphics[width=7cm]{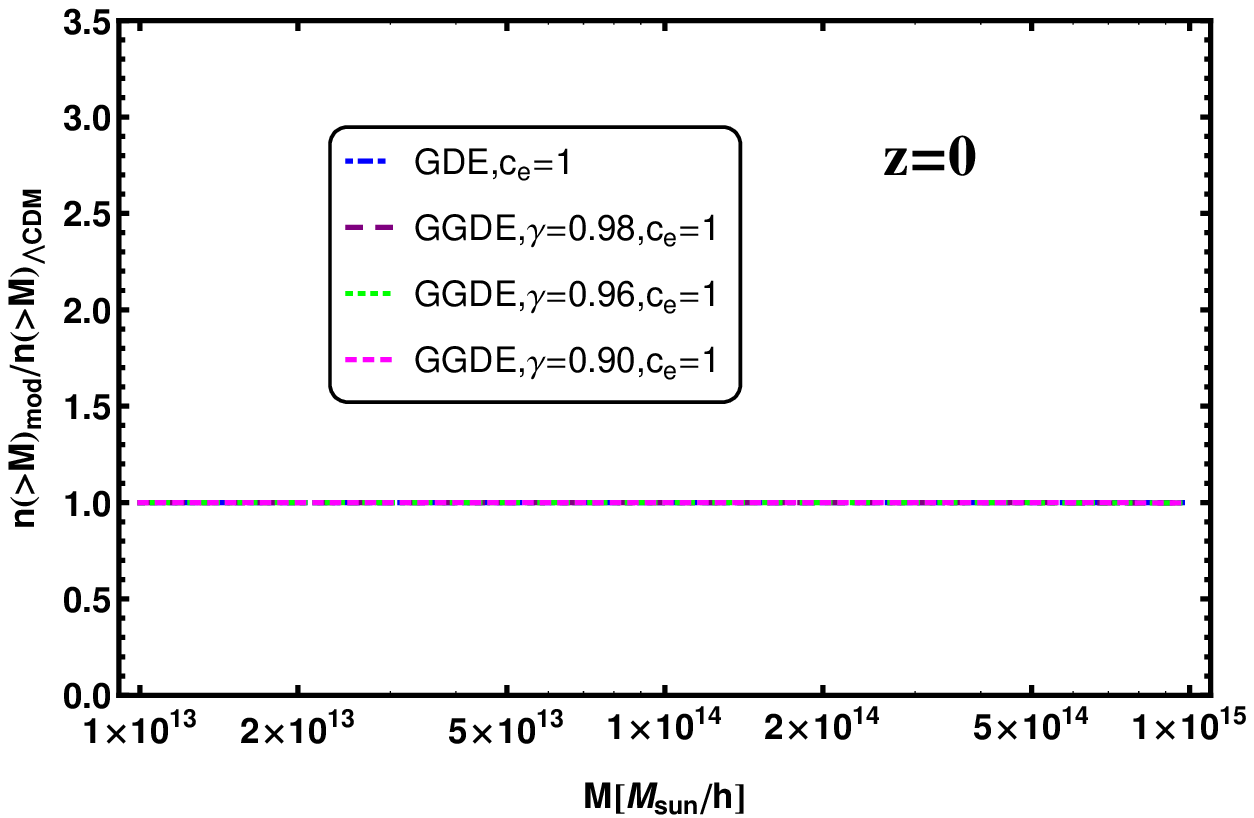}
  \includegraphics[width=7cm]{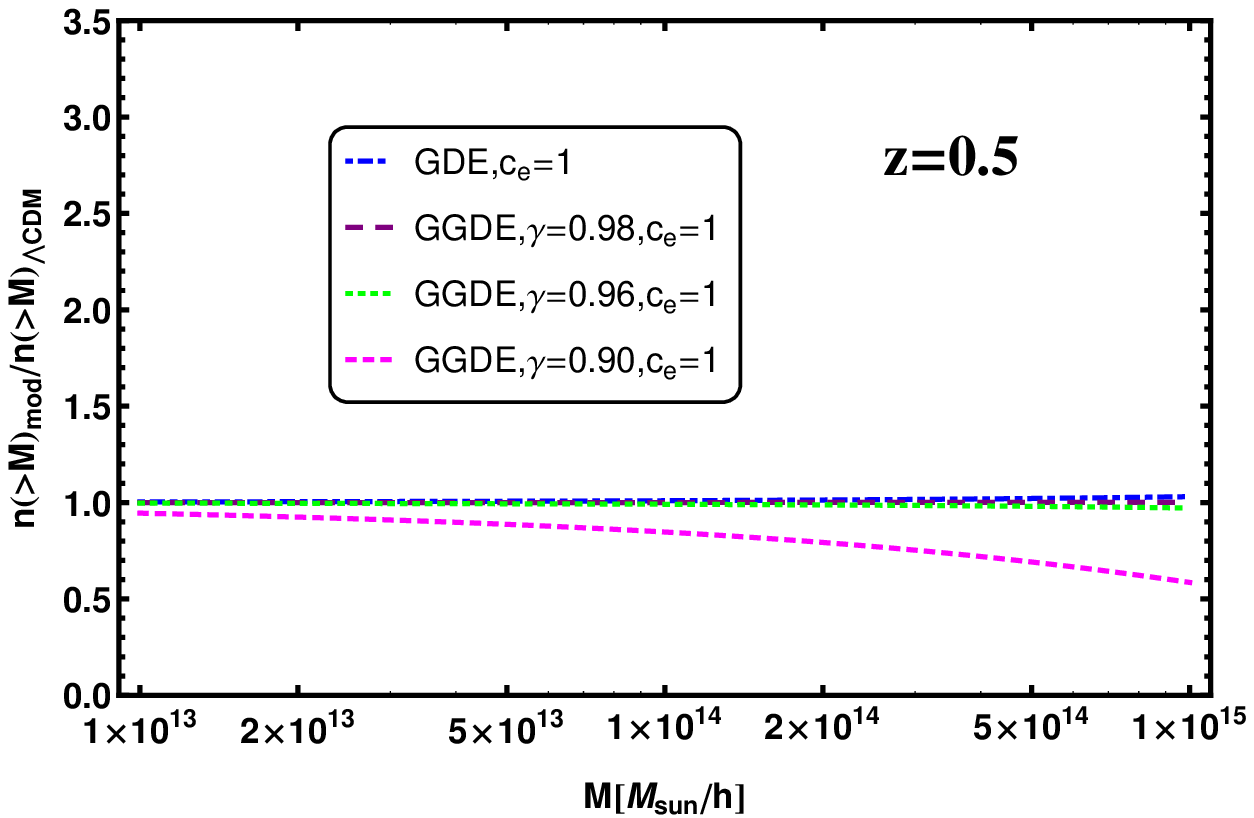}
  \includegraphics[width=7cm]{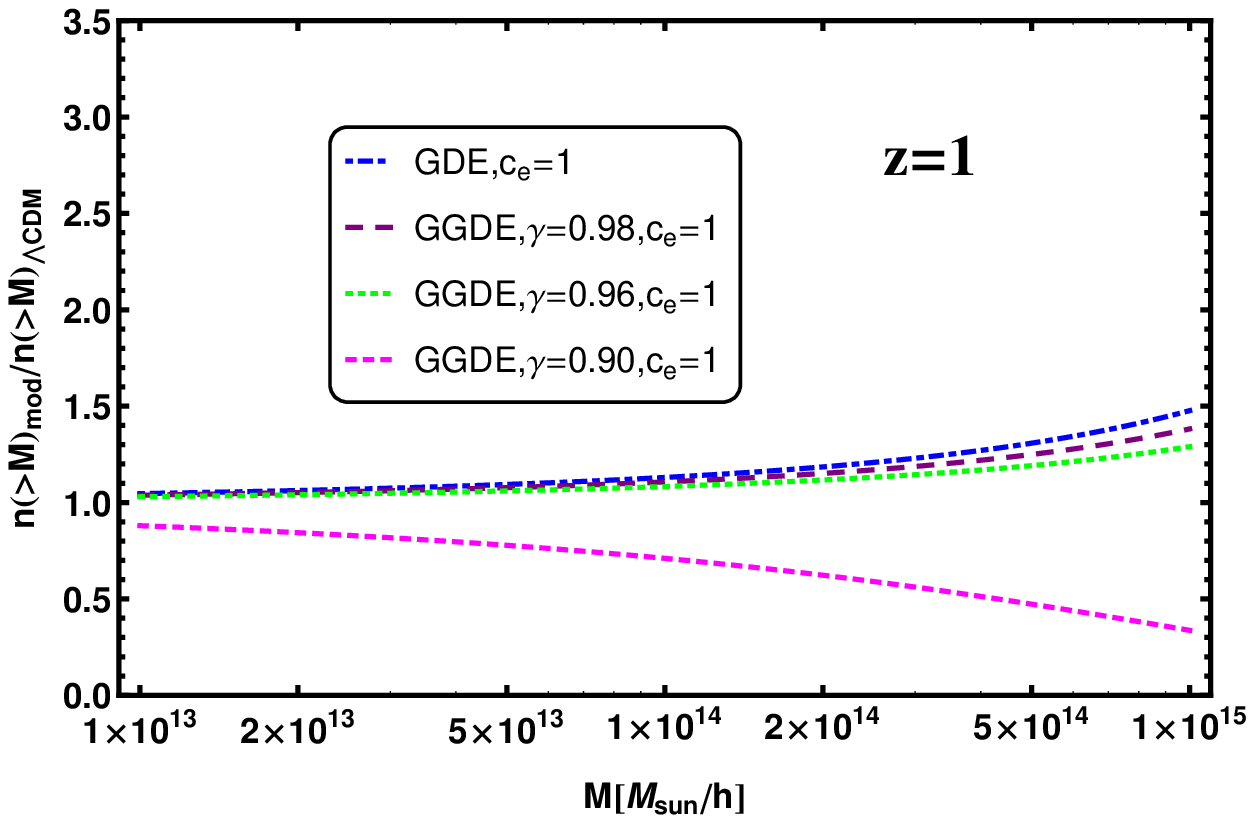}
  \includegraphics[width=7cm]{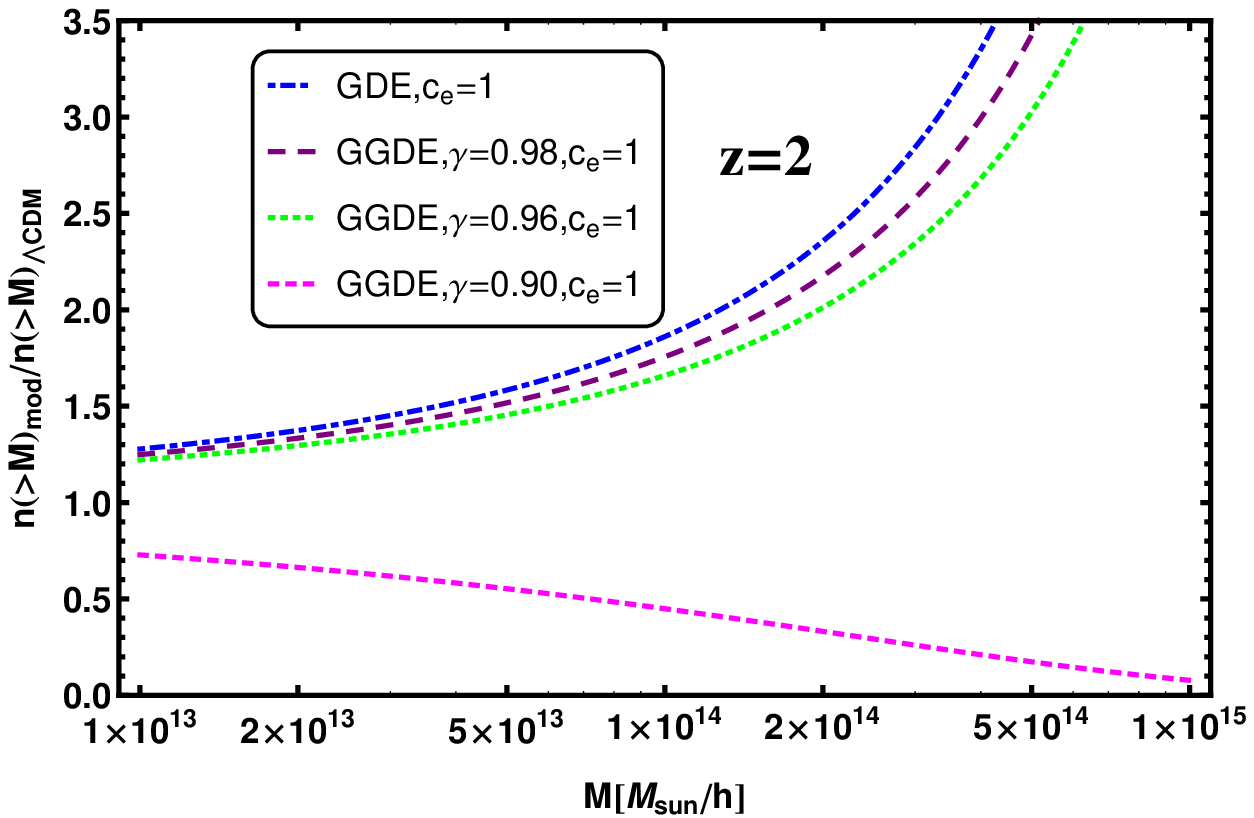}
  \caption{Ratio of the number of objects above a given mass $M$ for halos at $z=0$ (top left), $z=0.5$ (top right),
  $z=1.0$ (bottom left) and $z=2.0$ (bottom right) between the ghost and generalised ghost DE models and the
  concordance $\Lambda$CDM model. The blue dotted-dashed curve represents the ghost DE model. The purple dashed,
  green dotted and pink dashed curves stand for generalised ghost DE model with model parameter $\gamma=0.98$,
  $\gamma=0.96$ and $\gamma=0.90$, respectively.}
  \label{fig:mass_fun1}
 \end{center}
\end{figure*}

In this section, having at hands all the necessary ingredients, we can compute the predicted number density of
virialised objects in the Press-Schechter formalism for ghost and generalised ghost DE cosmologies. As first step,
we use equation (\ref{eq:ndensity}) for non-clustering ghost and generalised ghost DE models. In next section we
investigate the predicted number density of virialised clusters on the basis of the formulation presented in
\cite{Creminelli2010,Basse2011,Batista2013} where the total mass of the halos is affected by DE perturbations.

We use the $f_{\rm ST}$ mass function given in equation
(\ref{eq:multiplicity_ST}) and choose four different redshifts:
$z=0$, $z=0.5$, $z=1.0$ and $z=2.0$. In order to calculate
$\sigma^2$, we follow the formulations presented in
\cite{Abramo2007} and \cite{Naderi2015}. We also adopt the
$\Lambda$CDM model as reference model with normalization of the
matter power spectrum $\sigma_{\rm 8} = 0.776$, in agreement with
recent observations \citep{Planck2014_XVI,Planck2014_XX}. In figure
(\ref{fig:mass_fun1}), we show the ratio of the number of objects
above a given mass $M$ between the ghost and generalised ghost DE
models and the concordance $\Lambda$CDM model, in the case of
non-clustering DE cosmologies. In the case of generalised ghost DE
model, we choose three different values for the model parameter:
$\gamma=0.90$, $\gamma=0.96$ and $\gamma=0.98$. As was discussed in
section~(\ref{sect:GDE}), for $\gamma=1.0$, the generalised ghost
model reduces to the ghost DE model. In the upper left (right)
panel, the results are presented for redshift $z=0$ ($z=0.5$). The
lower left (right) panel is for halos at redshift $z=1$ ($z=2$), as
indicated in the legends. Due to the identical normalization of the
matter power spectrum, we see that at $z=0$ all models have the same
number of objects. At $z=0.5$, the ghost ($\gamma=1$) and
generalised ghost with $\gamma=0.96$ and $\gamma=0.98$ are still
giving the same number of object compared to the concordance
$\Lambda$CDM model, with very small differences for the large mass
tail of the distribution ($10^{15}M_{\odot}/h$). However, for
smaller values of $\gamma$ the differences between the generalised
ghost and the $\Lambda$CDM model are considerable. For
$\gamma=0.90$, the predicted number density for virialised halos at
high mass tail ($10^{15}M_{\odot}/h$) is roughly $40\%$ lower than
the concordance $\Lambda$CDM model (see top-right panel of
figure~(\ref{fig:mass_fun1}). At higher redshifts, $z=1$ and $z=2$,
the predicted number of structures exceeds what predicted by the
$\Lambda$CDM model, in the cases of ghost and generalised ghost DE
models with model parameter $\gamma$ close to unity, i.e.,
$\gamma=0.96, 0.98$. The generalised ghost model with low value
$\gamma=0.90$ shows a decrement in the number of objects compared to
other cases. We see that a major difference between ghost DE models
and $\Lambda$CDM model takes place at high mass while all models are
roughly producing the same number of objects at the low mass tail,
as expected. From a more quantitative point of view, at redshift
$z=1$, the ghost DE model shows an increment in the number of
structures of $\approx 47\%$, while for the generalised ghost DE
model with $\gamma=0.98$ ($\gamma=0.96$) the increment is $\approx
38\%$ ($\approx 29\%$), when they are compared to the standard
$\Lambda$CDM model. Surprisingly, the case with $\gamma=0.90$ shows
a substantial lack of high mass objects, of the order of $\approx
66\%$ for objects of mass $\approx 10^{15}M_{\odot}/h$.

\subsection{Corrected mass function in clustering ghost DE models}
It is well known that in EdS cosmology $y=R_{\rm vir}/R_{\rm
ta}=1/2$. In the EdS model, the virial overdensity can be calculated
analytically and the calculations lead to $\Delta_{\rm vir}\simeq
178$ independently of the redshift $z$. It is also well known that
this value strongly depends on the particular background cosmology
and changes in the presence of DE. It was shown that the
virialization process of dark matter overdensities in the non-linear
regime depends on the properties of DE models
\citep{Lahav1991,Maor2005,Creminelli2010,Basse2011}. Moreover, in
clustering DE models, we should take into account the contribution
of DE perturbations to the total mass of the halos
\citep{Creminelli2010,Basse2011,Batista2013,Pace2014a}. Depending on
the quintessence or phantom DE EoS parameter, $w_{\rm de}(z)$, DE
can add or subtract mass to the total mass of the halo,
respectively. The fraction of DE mass to be taken into account with
respect to the mass of the dark matter is given by the quantity
$\epsilon(z)=M_{\rm de}/M_{\rm m}$. The mass of dark matter $M_{\rm
m}$ is defined as
\begin{equation}
M_{\rm m}=4\pi\bar{\rho}_{\rm m}\int_0^{\rm R_{vir}}dR~R^2(1+\delta_{\rm m})\;.
\end{equation}
where $\bar{\rho}_{\rm m}$ is the mean density of dark matter. The mass of DE in the virialization process
depends on what it is considered to be the mass of the DE component. If we only assume the contribution of DE
perturbation, then the mass of the DE component is given by
\begin{equation}\label{eq:DEmass1}
M_{\rm de}^{\rm P}=4\pi\bar{\rho}_{\rm de}\int_0^{\rm R_{vir}}dR R^2 \delta_{\rm de}(1+3c_{\rm e}^2)\;,
\end{equation}
where $\bar{\rho}_{\rm de}$ is the mean density of dark energy. On the other hand, if we assume also the contribution
at the background level (in analogy to dark matter), the total mass of DE is given by
\begin{equation}\label{eq:DEmass2}
M_{\rm de}^{\rm T}=4\pi \bar{\rho}_{\rm de} \int_0^{\rm R_{vir}}dR R^2
\left[(1+3w_{\rm de})+\delta_{\rm de}(1+3c_{\rm e}^2)\right]\;.
\end{equation}

\begin{figure}
 \centering
 \includegraphics[width=8cm]{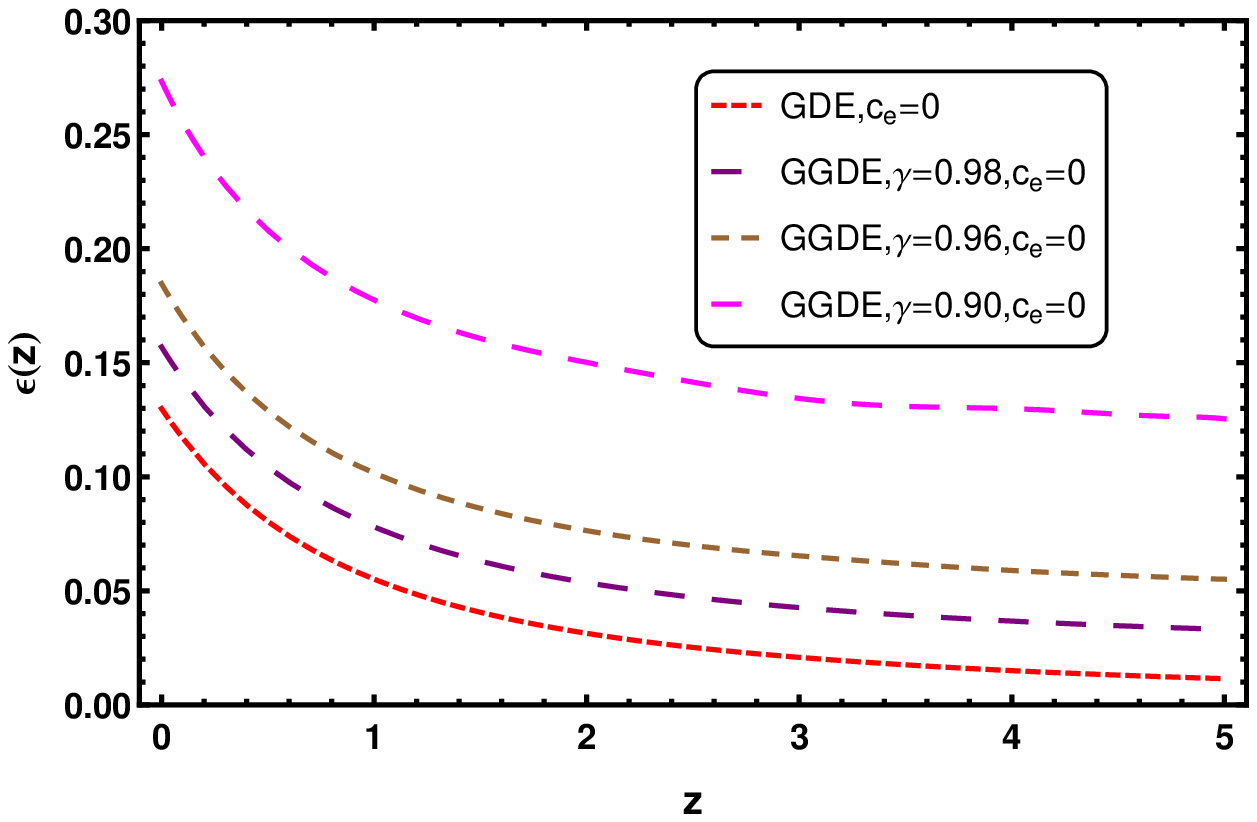}
 \includegraphics[width=8cm]{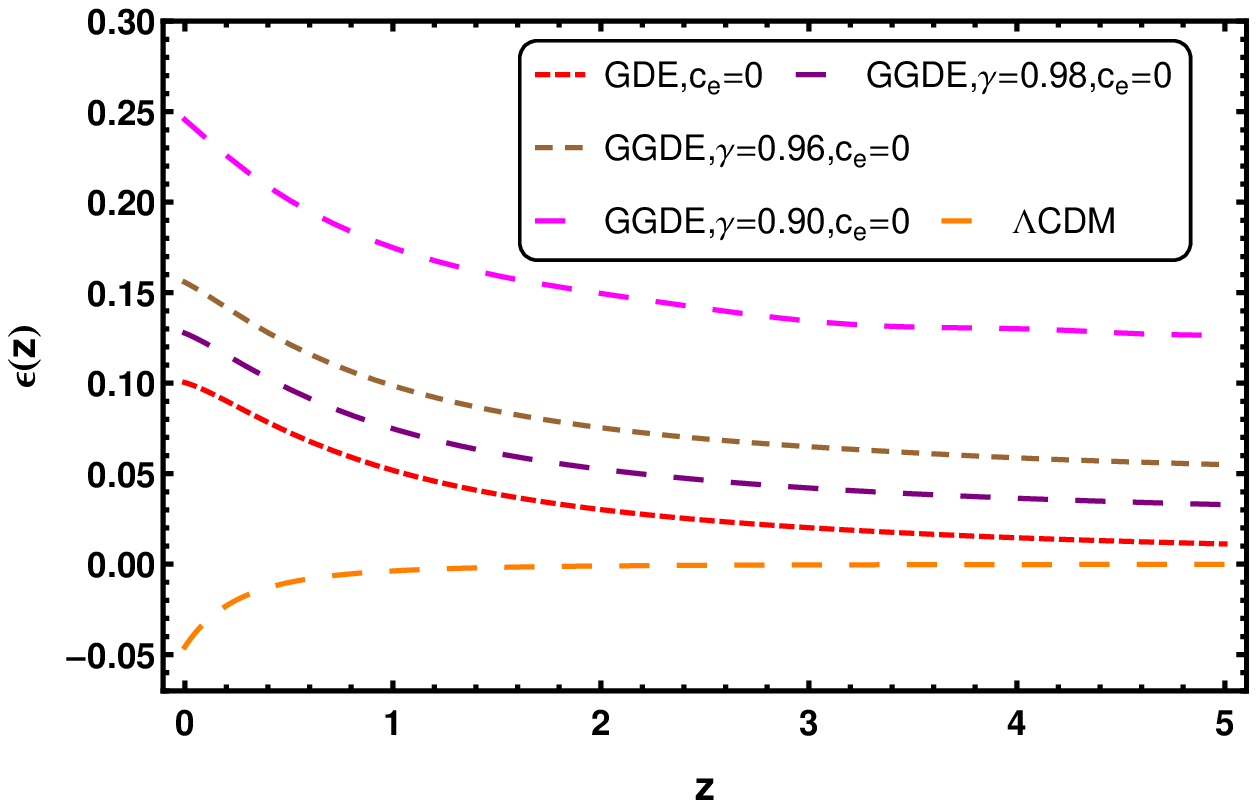}
 \caption{The ratio of DE mass to dark matter mass by using the definition of $\epsilon$ via equation
  (\ref{eq:epsilon1}) in the top panel and equation (\ref{eq:epsilon2}) in the bottom panel. The red dotted-dashed
  curve represents the clustering ghost DE model. The purple long dashed, brown dashed and pink long-dashed curves
  stand for clustering generalised ghost DE model with model parameter $\gamma=0.98$, $\gamma=0.96$ and $\gamma=0.90$,
  respectively. The orange long dashed one indicates the $\Lambda$CDM model.}
 \label{fig:epsilon}
\end{figure}

Here we compute the quantity $\epsilon(z)$ in the case of full
clustering DE scenario. It should be noted that due to the
background contribution in equation (\ref{eq:DEmass2}), even in the
case of homogeneous DE scenarios, DE can add or subtract mass to the
dark matter halos \citep[see also][]{Batista2013,Pace2014a}. Since
we work in the framework of the top-hat spherical profile, the
quantities inside the collapsing sphere evolve only in time without
any spatial dependency. Hence on the basis of the definition in
equation (\ref{eq:DEmass1}) we have
\begin{equation}\label{eq:epsilon1}
\epsilon(z)=\frac{\Omega_{\rm de}(z)}{\Omega_{\rm m}(z)}\frac{\delta_{\rm de}}{1+\delta_{\rm m}}\;,
\end{equation}
and under the same assumption for equation (\ref{eq:DEmass2})
\begin{equation}\label{eq:epsilon2}
\epsilon(z)=\frac{\Omega_{\rm de}(z)}{\Omega_{\rm m}(z)}\frac{1+3w_{\rm de}(z)+\delta_{\rm de}}{1+\delta_{\rm m}}\;.
\end{equation}

In figure~(\ref{fig:epsilon}) we show the evolution of $\epsilon(z)$ according to equations~(\ref{eq:epsilon1}) (top
panel) and~(\ref{eq:epsilon2}) (bottom panel) for both ghost and generalised ghost DE models. We refer the reader to
the caption for line styles and colours. We notice that according to equation (\ref{eq:epsilon1}), both ghost and
generalised ghost DE models give a positive contribution to the total mass of the halos ($\epsilon(z)>0$). The
quantity $\epsilon$ grows with redshift $z$ meaning that the contribution of DE on the total mass is larger for halos
which virialize at lower redshifts. This is explained by taking into account the fact that at late times dark energy
perturbations arise, while being negligible at earlier times. In particular, the generalised ghost DE model gives a
higher contribution of DE to the total mass of halos compared to ghost DE model. We also see that a lower value of the
model parameter $\gamma$ results in a higher value of $\epsilon (z)$. On the other hand, using the definition in
equation~(\ref{eq:epsilon2}), the behaviour of $\epsilon$ is different. Looking at equation~(\ref{eq:epsilon2}), we
see that the sign of the quantity $1+3w_{\rm de}+\delta_{\rm de}$ determines how DE contributes to the total mass of
the halo. During the history of the halo, when $3w_{\rm de}+\delta_{\rm de}<-1$, the contribution of the DE mass is
negative and therefore it lowers the total mass of the halos. The other important point to note is that according to
the definition of $\epsilon$ via equation~(\ref{eq:epsilon2}), one can consider the DE mass even for homogeneous DE
models ($\delta_{\rm de}=0$). Hence in the bottom panel of figure~(\ref{fig:epsilon}) we show $\epsilon$ for the
$\Lambda$CDM model ($w_{\rm de}=-1$). We also see that for the $\Lambda$CDM model $\epsilon$ is always negative. On
the other hand, when $3w_{\rm de}+\delta_{\rm de}>-1$, the contribution of DE is positive. This is the behaviour
taking place for ghost and generalised ghost DE models. Comparing the top and bottom panels in
figure~(\ref{fig:epsilon}), we see that for ghost and generalised DE models the values of $\epsilon$ according to the
second definition are lower than equation~(\ref{eq:epsilon1}).

\begin{figure*}
 \begin{center}
  \includegraphics[width=7cm]{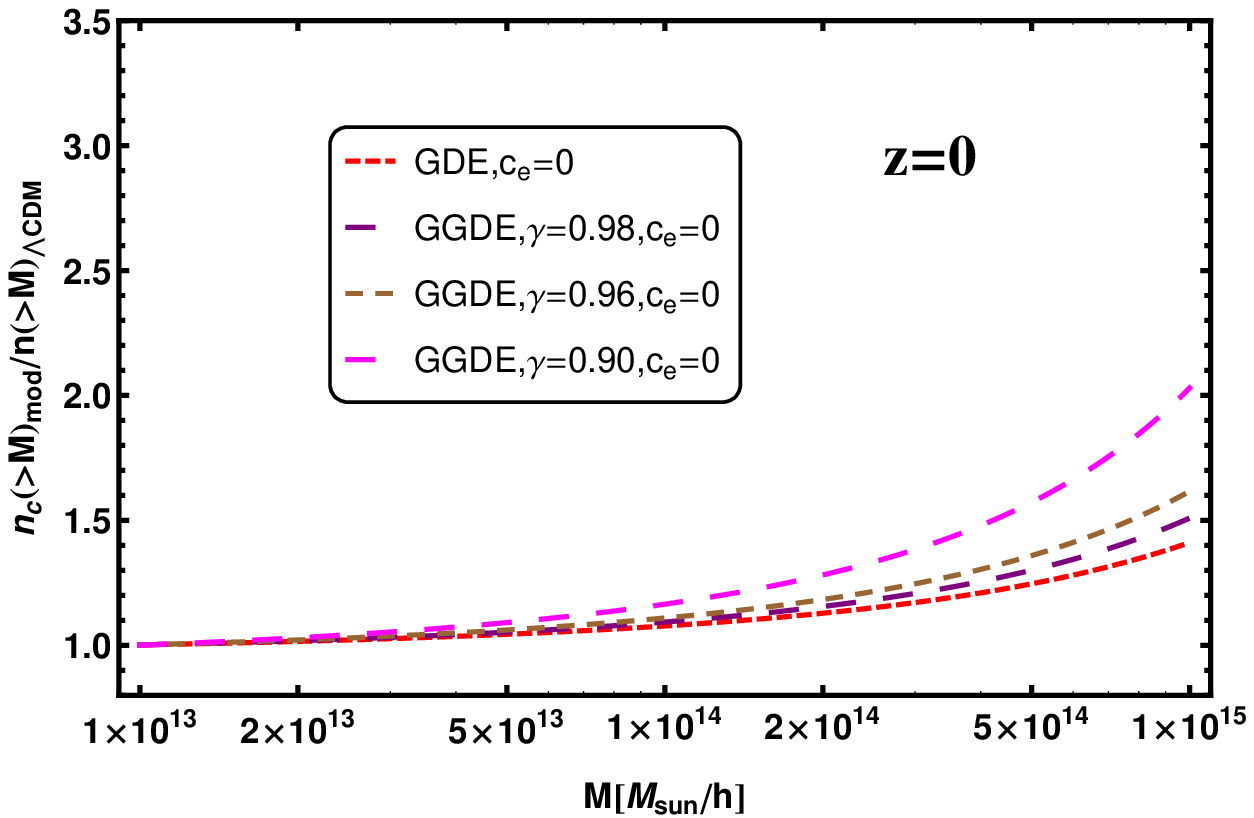}
  \includegraphics[width=7cm]{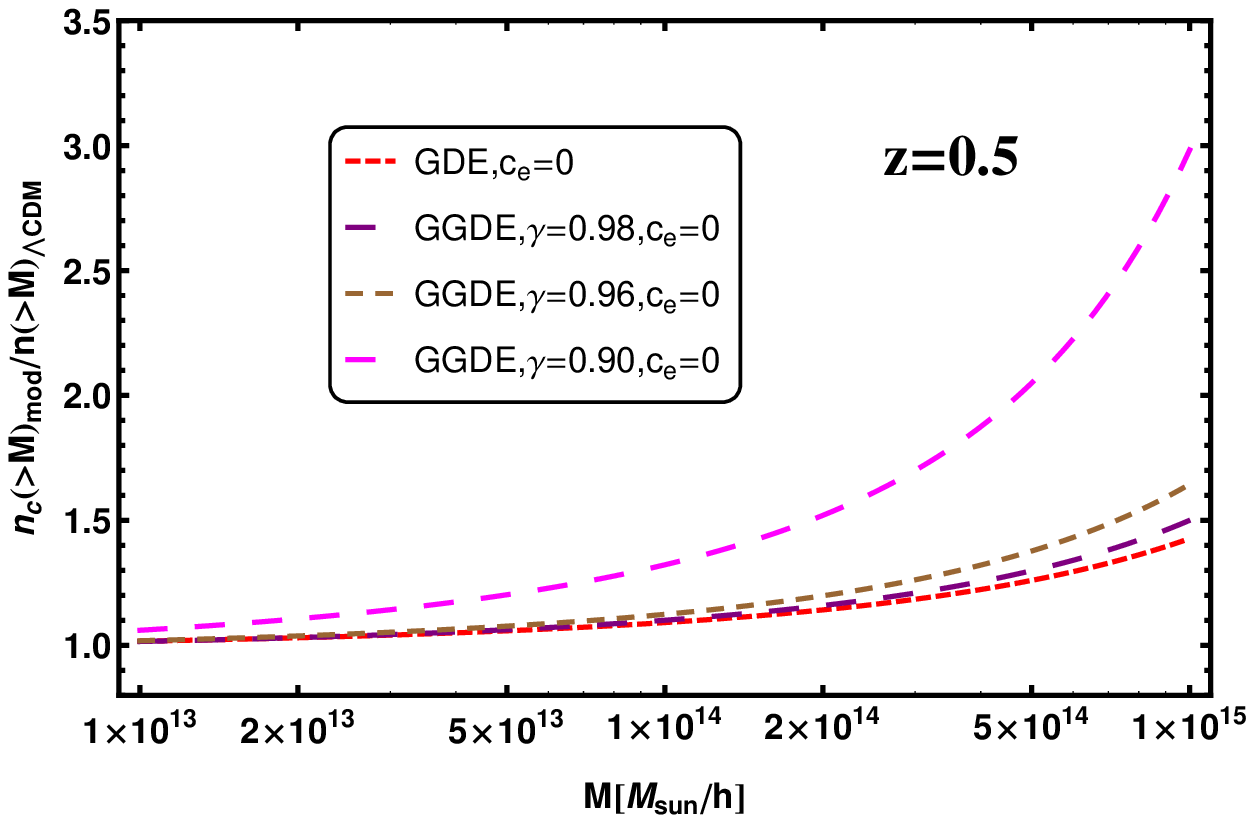}
  \includegraphics[width=7cm]{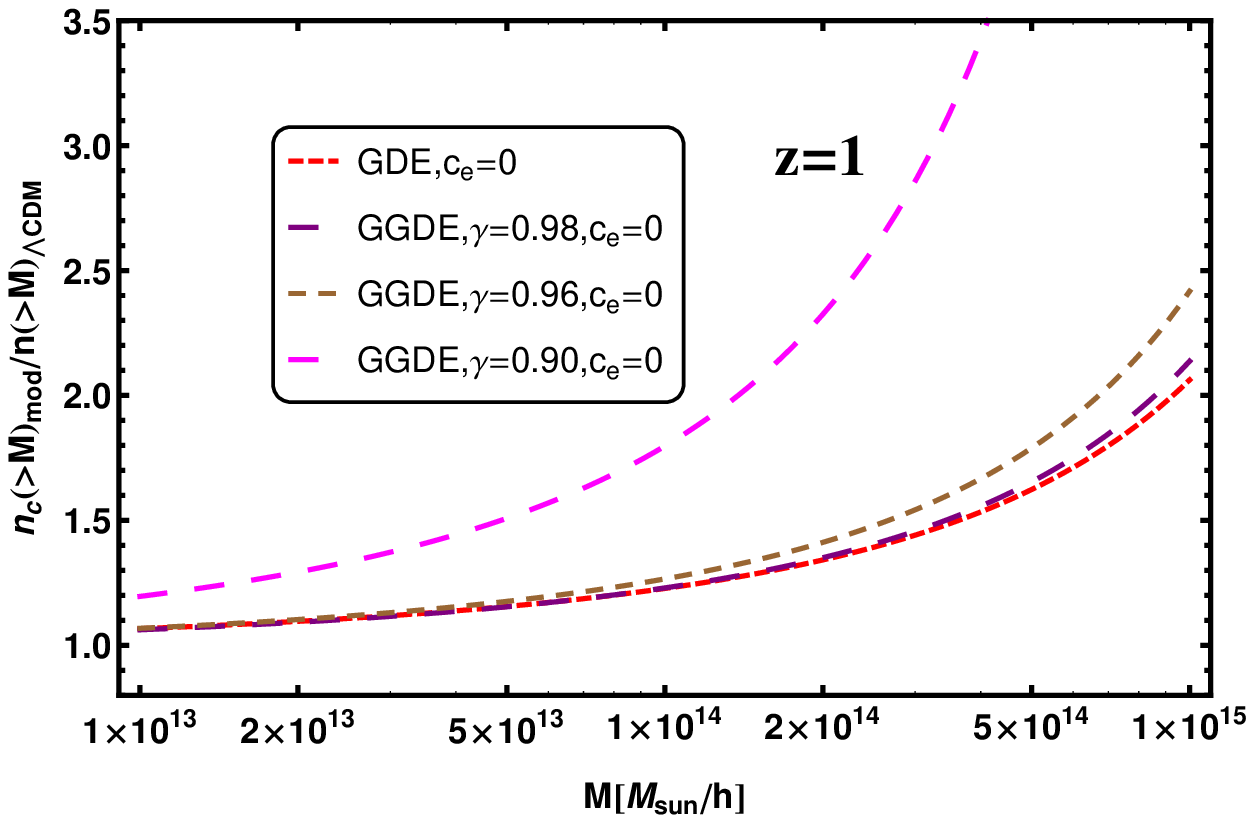}
  \includegraphics[width=7cm]{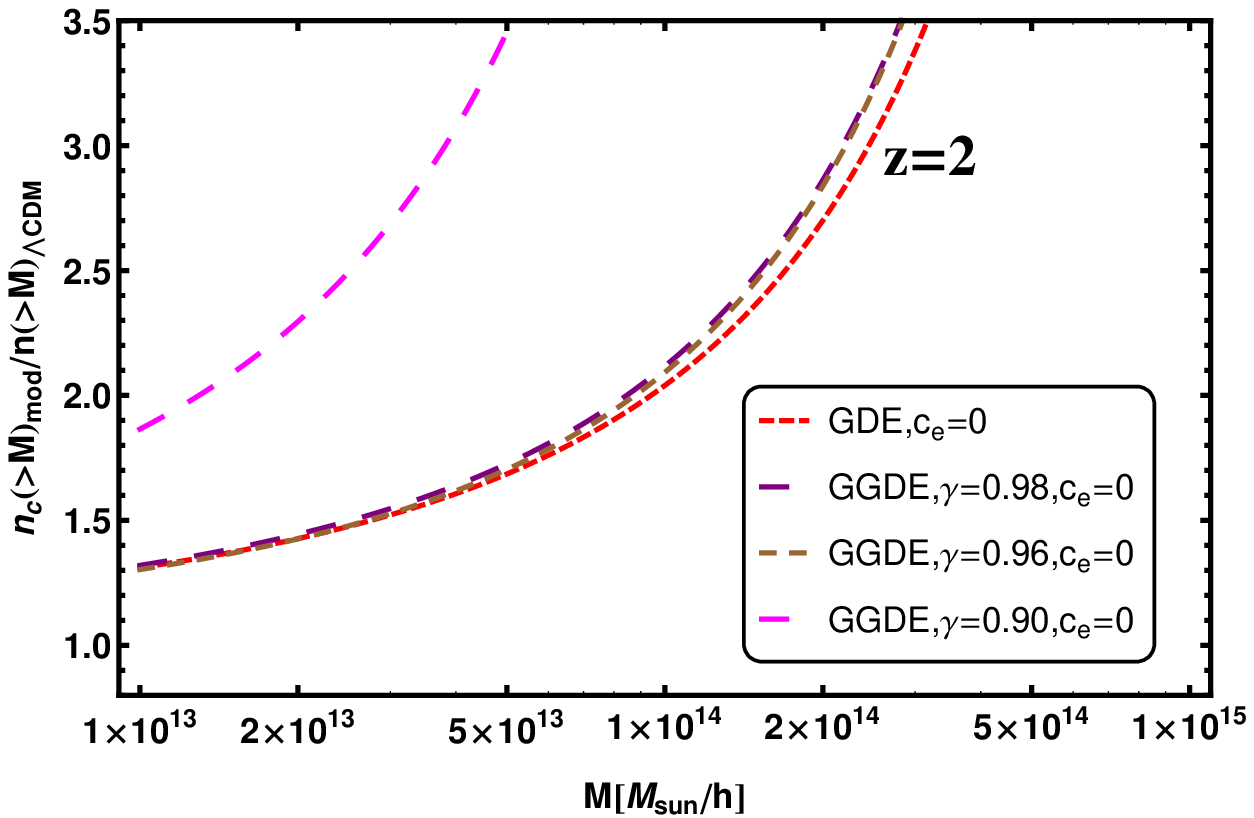}
  \caption{Ratio of the number of objects above a given mass $M$ for halos at $z=0$ (top left panel), $z=0.5$ (top
  right panel), $z=1.0$ (bottom left panel) and $z=2.0$ (bottom right panel) between the ghost and generalised ghost
  DE models and the concordance $\Lambda$CDM model by using the mass definition in equation (\ref{eq:epsilon1}). The
  red dotted-dashed curve represents the clustering ghost DE model. The purple long dashed, brown dashed and pink
  long-dashed curves stand for clustering generalised ghost DE model with model parameter $\gamma=0.98$, $\gamma=0.96$
  and $\gamma=0.90$, respectively.}
  \label{fig:mass_fun2}
 \end{center}
\end{figure*}

We now compute the number density of virialized halos in the
presence of DE mass correction. Following the procedure outlined in
\cite{Batista2013} and \cite{Pace2014a}, in the presence of DE
perturbations, we change the mass of halos as $M\rightarrow
M(1-\epsilon)$. In this case the corrected mass function can be
redefined as follows \citep{Batista2013}
\begin{equation}
\frac{dn_{\rm c}(z,M)}{dM}=-\frac{\rho_{\rm m0}}
{M(1-\epsilon)}\frac{d\ln{\sigma(M,z)}}{dM}f(\sigma)\;,\label{eqn:PScorrected}
\end{equation}
where $f(\sigma)$ is given by equation (\ref{eq:multiplicity_ST}). It should be noted that the clustering of the DE
component can also change the mass function $f(\sigma)$ by changing the quantities $\delta_{\rm c}$ and $\sigma(M,z)$.
In figure (\ref{fig:mass_fun2}) we show the relative number density by using the corrected mass function formula in
equation (\ref{eqn:PScorrected}) and the definition of $\epsilon$ in equation (\ref{eq:epsilon1}). Analogously to
figure (\ref{fig:mass_fun1}), we choose four different redshifts $z=0, 0.5, 1$ and $z=2$. We see that due to the
contribution of the DE mass, the relative number density between ghost and generalised DE models and $\Lambda$CDM
model $n_{\rm c}/n_{\rm \Lambda CDM}$ is equal or larger than one. In the top panels, for $z=0$ and $z=0.5$, all
ghost, generalised ghost and $\Lambda$CDM models generate the same number of objects at the low-mass end. However, at
high masses, generalised ghost DE models with smaller value of the model parameter $\gamma$ produce more objects. At
higher redshifts, $z=1$ and $z=2$, one can see that ghost and generalised ghost DE models have more objects for both
the low mass and the high mass tail compared to the concordance $\Lambda$CDM model. Comparing figures
(\ref{fig:mass_fun1}) and (\ref{fig:mass_fun2}), we conclude that for all the redshifts here considered, the predicted
relative number density of halos in the presence of DE corrections to the halo mass is higher than what was found for
homogeneous DE scenarios. In particular, the differences are larger for high-mass objects. This feature of ghost and
generalised ghost DE models is in agreement with the results of \cite{Batista2013} for inhomogeneous Early Dark
Energy models. To compare the models quantitatively, we restrict our analysis to $z=0$. While non-clustering DE
models and the $\Lambda$CDM model have the same number of objects (top-left panel of figure~(\ref{fig:mass_fun1})),
clustering DE models deviate from the $\Lambda$CDM model particulary at the high-mass tail (top-left panel of
figure~(\ref{fig:mass_fun2})). We see that clustering generalised ghost DE model with $\gamma=0.90$ doubles the values
obtained for the $\Lambda$CDM model, while for $\gamma=0.96$ the increment in the number of objects is $\approx 61\%$.
Finally, for $\gamma=0.98$, differences are of the order of $50\%$. As expected, the increase in $\gamma$ corresponds
to a decrease in the differences with respect to the $\Lambda$CDM model at the high mass tail (of the order of $40\%$
for the ghost dark energy model).

\section{Conclusions}\label{sect:conclusions}
In this work we studied the non-linear evolution of structure formation in ghost and generalised ghost dark energy
models within the framework of the spherical collapse model. Dark energy models are described by a constant or a time
varying equation of state $w(a)$ whose functional form is usually purely phenomenological. The advantage of dynamical
dark energy models with respect to the $\Lambda$CDM model is that to alleviate its theoretical problems. Another
interesting aspect is the presence of fluctuations in the dark energy fluid that can have an important and
characteristic impact on non-linear structure formation. Ghost dark energy models are theoretically well motivated and
trace their origin back to the studies of the low energy effective QCD theory
\citep{Veneziano1979,Witten1979,Kawarabayashi1980,Rosenzweig1980}. Here we concentrate on the cosmological
implications of such models and present results that can be useful from an observational point of view.

We start our analysis by studying the effects of the modification of the background expansion history of the growth
factor and on a second step we take into account also the perturbations in the dark energy component. We showed that
the linear growth factor is sensitive to the details of the model considered. In particular, when dark energy is
homogeneous, differences with respect to the $\Lambda$CDM model are very pronounced and can be up to 50\% at
relatively high redshifts. Interestingly, when dark energy fluctuations are taken into account, differences become
smaller and the models studied get closer to the $\Lambda$CDM model. This is exactly what happens for early dark
energy model and it applies to this class of models as well, being the amount of dark energy at early times
significantly more important (see bottom panel of figure~(\ref{fig:back})).

Being the equations governing the evolution of the linearly
extrapolated overdensity parameter $\delta_{\rm c}$ identical to
those of the linear growth factor, results are similar to what found
for the linear growth factor. Once again when the perturbations in
the dark energy fluid are included in the calculations, the models
examined resemble closer the $\Lambda$CDM model. This is easily
explained by taking into account that when dark energy perturbations
play a role, Poisson equation is modified and this makes the model
more similar to the $\Lambda$CDM one \citep[see][for a more detailed
discussion]{Batista2013}. Similar conclusions can be drawn for the
non-linear virial overdensity $\Delta_{\rm V}$.

While the growth factor can be inferred from the determination of the matter power spectrum on linear scales at a
given redshift or by measuring the quantity $f\sigma_8(z)$, where $f=d\ln{D_{+}}/d\ln{a}$ and
$\sigma_8(z)=D_{+}(z)\sigma_8(z=0)$, the linear extrapolated overdensity $\delta_{\rm c}$ is a purely theoretical
quantity and it would be very hard to infer it from observations. Nevertheless they can be combined together and be
the basic ingredients to evaluate the mass function and the number of halos above a given mass. When keeping the dark
energy fluid smooth and important only at the background level, we showed that both the ghost and generalised ghost
models give an excess of objects with respect to the $\Lambda$CDM model, except for the generalised ghost model with lower value
$\gamma=0.9$. As expected, differences are very pronounced only for the high mass tail of the mass function, while at
low masses the models are practically indistinguishable.

One might consider that if dark energy can clump, then the total mass of the halos should be affected. By defining the
contribution of dark energy limited to that of the clumps, we showed in the top panel of figure~(\ref{fig:epsilon})
that the dark energy mass can be at least 10\% of the dark matter mass at $z=0$, with differences growing with
decreasing the parameter $\gamma$ and decreasing at high redshift where the contribution of dark energy is less
important. Note that if the amount of dark energy at early times is significantly bigger than in the $\Lambda$CDM
model, then the dark energy mass can still be an appreciable fraction of the dark matter mass. By taking into account
this correction and evaluating the number of objects above a given mass, we showed that the dark energy models we
studied in this work predict much more objects with respect to the $\Lambda$CDM model at all redshifts. This could be a
good test to compare theoretical predictions with observations.

{\footnotesize
\bibliographystyle{mn2e}
\bibliography{ref}
}

\bsp

\label{lastpage}

\end{document}